\documentclass[10pt,USletter,twocolumn]{article}
\usepackage{amsmath}
\usepackage{amsfonts}
\usepackage{amssymb}
\usepackage{graphicx}
\usepackage[left=1.5cm,right=1.5cm,top=2cm,bottom=2cm]{geometry}

\usepackage{setspace}

\usepackage{cite}
\usepackage[electronic]{ifsym}
\usepackage{indentfirst,latexsym}
\usepackage{bm}
\usepackage{bbm}
\usepackage{amsmath,amsfonts,amsthm,amscd,amssymb}
\usepackage{pifont}
\usepackage[mathscr]{euscript}
\usepackage{url}

\usepackage{graphicx}
\usepackage{subfigure}

\usepackage[all,pdf]{xy}

\usepackage{booktabs}
\usepackage{multirow}
\usepackage{stfloats}

\DeclareMathOperator{\dif}{d}  
\DeclareMathOperator{\timer}{\mathscr{T}}

\newcommand{\tic}[2]{\timer_{#1}{#2}}

\renewcommand{\vec}[1]{\bm{#1}}

\newcommand{\mfloor}[1]{ \left\lfloor {#1} \right\rfloor }

\newcommand{\mat}[1]{\bm{#1}}

\newcommand{\mj} {\mathrm{j}}

\newcommand{\me} {\mathrm{e}}

\newcommand{\fracode}[2]{\frac{\dif {#1}}{\dif {#2}}}         
\newcommand{\fracpde}[2]{\frac{\partial {#1}}{\partial {#2}}} 

\newcommand{\set}[1]{\left\{ #1 \right\}}
\newcommand{\seq}[1]{\langle #1 \rangle}
\newcommand{\abs}[1]{\left| #1 \right|}

\newcommand{\braket}[2]{ \langle #1 | #2 \rangle}
\newcommand{\norm}[1]{\left\lVert #1 \right\rVert}

\newcommand{\trsp}[1]{{#1}^\textsf{T}}

\newcommand{\inv}[1]{#1^{-1}}

\newcommand{\quat}[1]{\mathbbm{#1}}       

\newcommand{\ES}[3]{\mathbb{#1}^{{#2}\times {#3}}}     
\newcommand{\GrGL}[2]{\mathrm{GL}(#1,\mathbb{#2})}
\newcommand{\scru}[2]{{#1}^{\mathrm{#2}}}
\newcommand{\scrd}[2]{{#1}_{\mathrm{#2}}}

\usepackage{cleveref}
\usepackage{algorithm}         
\usepackage{algorithmicx}
\usepackage{algpseudocode}
\usepackage{float}

\usepackage{forest}

\usepackage{longtable}
\usepackage{supertabular}
\usepackage{caption}

\newcommand{\Algr}{\textbf{Algorithm}~}
\newcommand{\Fig}{\textbf{Figure}~}
\newcommand{\Tab}{\textbf{Table}~}

\algnewcommand\algorithmicswitch{\textbf{switch}}
\algnewcommand\algorithmiccase{\textbf{case}}
\algnewcommand\algorithmicdefault{\textbf{default}}
\algnewcommand\algorithmicassert{\cpvar{assert}}
\algnewcommand\Assert[1]{\State \algorithmicassert(#1)}%
\algdef{SE}[SWITCH]{Switch}{EndSwitch}[1]{\algorithmicswitch\ #1\ \algorithmicdo}{\algorithmicend\ \algorithmicswitch}%
\algdef{SE}[CASE]{Case}{EndCase}[1]{\algorithmiccase\ #1}{\algorithmicend\ \algorithmiccase}%
\algdef{SE}[DEFAULT]{Default}{EndDefault}[1]{\algorithmicdefault\ #1}{\algorithmicend\ \algorithmicdefault}%

\makeatletter
\renewcommand{\ALG@name}{Algorithm}
\newenvironment{breakablealgorithm}
{
	\begin{center}
		\refstepcounter{algorithm}
		\setlength{\baselineskip}{15pt} 
		\renewcommand{\caption}[2][\relax]{
			\hrule height.9pt depth0pt \kern0pt
			{\raggedright\textbf{\ALG@name~\thealgorithm} ##2\par}%
			\ifx\relax##1\relax 
			\addcontentsline{loa}{algorithm}{
				\protect\numberline{\thealgorithm}##2}%
			\else 
			\addcontentsline{loa}{algorithm}{
				\protect\numberline{\thealgorithm}##1}%
			\fi
			\kern2pt\hrule\kern2pt
		}
	}{
		\kern3pt\hrule\relax
	\end{center}
}

\usepackage{caption}
\usepackage{listings}
\newcommand{\cpvar}[1]{\texttt{#1}}

\newtheorem{thm}{Theorem}

\newtheorem{lem}[thm]{Lemma}

\author{Hong-Yan Zhang\thanks{Corresponding author, e-mail: hongyan@hainnu.edu.cn}, Fei Liu, Yu Zhou and Man Liang\\ School of Information Science and Technology, Hainan Normal University, Haikou 571158, China}
\title{Even Order Explicit Symplectic Geometric Algorithms for \\Solving Quaternions in Guidance Navigation  and Control via\\ Diagonal Pad\'{e} Approximation and Cayley Transform\thanks{This work was supported in part by the Hainan Provincial Natural Science Foundation of China under Grant 2019RC199, in part by the specific research fund of The Innovation Platform for Academicians of Hainan Province, and in part by the National Natural Science Foundation of China under Grant 62167003.}}
\date{Aug. 7, 2024}

\begin{document}
\maketitle

\begin{abstract}
Quaternion kinematical differential equation (QKDE) plays a key role in navigation, control and guidance systems. Although explicit symplectic geometric algorithms (ESGA) for this problem are available, there is a lack of a unified way for constructing high order symplectic difference schemes with configurable order parameter and the fractional interval sampling problem should be treated carefully. We present even order explicit symplectic geometric algorithms to solve the QKDE with diagonal Pad\'{e} approximation via a four-step strategy. Firstly, the Pad\'{e}-Cayley lemma is proved  and used to simplify the symplectic Pad\'{e} approximation for the linear Hamiltonian system with infinitesimal symplectic structure. Secondly, both parallel and alternative iterative methods are proposed to construct the symplectic difference schemes with even order accuracy. Thirdly, the symplecity, orthogonality and invertibility of the single-step transition matrices are proved rigorously. Finally, the explicit symplectic geometric algorithms are designed for both the linear time-invariant and  linear time-varying QKDE. The maximum absolute error for solving the QKDE is $\mathcal{O}((t_f-t_0)\tau^{2\ell})$ where  $\tau$ is the time step, $\ell$ is the order parameter and $[t_0,t_f]$ is the time span. The linear time complexity and constant space complexity of computation  as well as the simple algorithmic structure show that our algorithms are appropriate for real-time applications in aeronautics, astronautics, robotics and so on. The performance of the proposed algorithms are verified and validated by mathematical analysis and numerical simulation.
\end{abstract}

\tableofcontents

\section{Introduction}  \label{sec-Intro}
Quaternions  have been widely utilized
in aerospace\cite{Wie1985,Wie1989,Wie1995,Wie2002,Kuipers2002}, aeronautics\cite{Friedland1978,Kim2004,Rogers2007,Zhong2012, Cooke1992,Allerton2009,Diston2009,JSBSim2014}, robotics  \cite{Yuan1988,Funda1990,Chou1992,Fjellstad1994,Zamani2013,Murartal2015orb-slam, CHRobotics}, computer vision\cite{Szeliski2010CVbook} and computer graphics\cite{Vince2011}. The comprehensive review of fundamental results about linear quaternion differential equations are presented by Kou \& Xia  \cite{Kou2018} and Kartiwa \cite{Kartiwa2023}.  The key advantages for quaternions stimulating many researches and applications lie in two facts:
\begin{itemize}
\item Geometrical singularity can be avoided when the rotation matrices are parameterized with quaternions instead of Euler angles.
\item Multiplications and additions are enough for  computing quaternions, which implies that quaternions are appropriate for real-time applications because of low computational complexity.
\end{itemize}

\noindent For the navigation, control and guidance  problems, the key issue is to find stable and precise numeric solution to the Robinson's \textit{quaternion kinematical differential equation} (QKDE) \cite{Robinson1958}
\begin{equation} \label{eq-QKDE}
\left\{
\begin{split}
&\fracode{\quat{q}}{t} = \frac{1}{2} \mat{\Omega}(\vec{\omega}(t)) \quat{q}, \quad  t\in [t_0, t_f] \\
&\quat{q}(t_0) = \quat{q}_0
\end{split}
\right.
\end{equation}
where $\quat{q}  = \trsp{[e_0, e_1, e_2, e_3]}$
is the vector representation of the quaternion $\quat{q}$, $[t_0, t_f]$ is the time span, $\quat{q}_0$ is the initial state, $\vec{\omega} =\trsp{[\omega_1(t), \omega_2(t), \omega_3(t)]}\in \ES{R}{3}{1}$
is the angular velocity\footnote{In references of aerospace and aeronautics, the notation $\vec{\omega} = \trsp{[p,q,r]}$ is used. In this paper, we use $p_i$ and $q_i$ to denote the $i$-th generalized momentum and displacement.} measured by the \textit{inertial measure unit} (IMU), and $\mat{\Omega} = \mat{\Omega}(\vec{\omega}(t))$ is specified by the $\vec{\omega}$ such that
\begin{equation} \label{eq-Omega}
\mat{\Omega}
      =-\trsp{\mat{\Omega}} = \begin{bmatrix}
       0 & -\omega_1 & -\omega_2  & -\omega_3 \\
       \omega_1 & 0 & \omega_3 & -\omega_2 \\
       \omega_2 & -\omega_3 & 0 & \omega_1 \\
       \omega_3 & \omega_2 & -\omega_1 & 0      
      \end{bmatrix}.
\end{equation}
If the $\vec{\omega}(t)$ is time independent, then the equation (\ref{eq-QKDE})  can be called \textit{autonomous} QKDE (A-QKDE) or \textit{linear time-invariant} QKDE (LTI-QKDE) since it is an autonomous system in the sense of classic mechanics and control theory and also it is a linear time-invariant (LTI) system in the sense of signals analysis.  On the contrary, if $\vec{\omega}(t)$ is time dependent,  the equation
(\ref{eq-QKDE}) can be named with the \textit{non-autonomous} QKDE (NA-QKDE) or \textit{linear time-varying} QKDE (LTV-QKDE) in general. 
Although the QKDE (\ref{eq-QKDE}) is a linear ordinary differential equation (ODE), it is challenging to find its solution for two reasons:
\begin{itemize}
\item The time-varying vector $\vec{\omega}(t)$ leads to a time-dependent matrix $\mat{\Omega}(\vec{\omega}(t))$. However, there is no general method for finding the \textit{analytical solution} (AS) to a LTV system.
\item The QKDE is critically (neutrally) stable and the numerical solution is sensitive to the computational errors \cite{Wie1985} because the eigenvalues of $\mat{\Omega}$ specified by \eqref{eq-Omega} contain two pure imaginary  numbers
$\pm \mj \norm{\vec{\omega}} = \pm \mj \sqrt{\omega_1^2 + \omega_2^2 + \omega_3^2}$ where $ \mj = \sqrt{-1}$.
\end{itemize}
Therefore, it is necessary to find a stable and precise integration method in the sense of long term time for solving  (\ref{eq-QKDE}). 

In the past decades, although there are  various different engineering-oriented \textit{single step} numeric approaches to solve the QKDE, the existing methods are inadequate in three perspectives:
\begin{itemize}
\item Serious accumulative computational errors are involved in the traditional non-symplectic methods.  The \textit{traditional finite difference scheme} (TFDS) is sensitive to the accumulative computational errors in the sense of long  term time. The typical methods include the   Taylor series method \cite{Hrastar1970, Cunningham1980, Wie1985}, classic Runge-Kutta method \cite{Cartwright1992RK-FDS1, Wolfram2024RK-FDS2, Press1992RK-FDS3} and the periodic normalization strategy \cite{Funda1990}. 
\item Earlier symplectic method with high computational complexity is not applicable in a real-time way. 
Compared with the TFDS, the  \textit{symplectic difference scheme} (SDS) can avoid accumulative computational errors automatically. Wang \cite{Wang2007} compared the Runge-Kutta scheme with SDS. The Gauss-Legendre difference scheme \cite{Butcher1964,Iserles1996}, an accurate and implicit Runge-Kutta method, is an \textit{implicit} SDS. However, its computational complexity is $\mathcal{O}(n^2)$ \cite{Varah1979} which means that it is  not appropriate for real-time applications. 
\item Low order non-configurable \textit{Explicit symplectic geometric algorithms} (ESGA) with fractional interval sampling  is not perfect for controlling moving object high speed and has extra difficulty in hardware implementation.   
In 2018, Zhang et al \cite{Zhang2018QKDE} proposed   two algorithms --- the ESGA-I and ESGA-II --- for the LTI-QKDE (A-QKDE) and LTV-QKDE (NA-QKDE) respectively,   which are not only precise in the sense of long term time, but also with linear time computational complexity and constant space computational complexity. However, there are some disadvantages for these two algorithms: 
\begin{itemize}
\item[i)] they  are not constructed with a unified way and can not be generalized for higher order cases for solving the QKDE with single step approach;
\item[ii)] the discrete time sequence corresponding to the interval $[t_0,t_f]$ is $\set{t_0, t_0+\frac{\tau}{2}, \cdots, t_0+\frac{(2k+1)\tau}{2}, \cdots}$ instead of $\set{t_0, t_0+\tau, \cdots, t_0+k\tau, \cdots}$ which implies that we have to treat the time intervals  carefully; 
\item[iii)] the order of the SDS's accuracy is not configurable and high order SDS is missing for moving objects with high speed such as
missiles and fourth-generation or fifth-generation fighters.
\end{itemize}
\end{itemize}
In consequence, it is necessary for us to explore reconfigurable high order ESGA which could avoid the dark sides i), ii) and iii) mentioned above.

It should be remarked that the accuracy of the measurement of angular velocity $\vec{\omega}(t)$ is of significant for the algorithms proposed for solving the LTI-QKDE and LTV-QKDE. In other words, the sensors for inertial measurement unit (IMU) is essential for practical applications. Recently, the quantum sensors have been invented for the IMU \cite{ENC2023-15437,qIMU2023,cqIMU2024}. It could be believed that the accuracy of measuring angular velocity will be wonderful in the near future. 

We remark that the theoretic exploring for solving the QKDE is still active in recent years. For example, Cai \& Kou proposed the two-sided coefficients method \cite{Cai2018}, Donachali \& Jafari \cite{Donachali2020} proposed the Adomian decomposition method, Liu et al. \cite{Liu2023} discussed QKDE in the perspective of solvability and Green's function. These discussions are not our emphasis since they are not engineering-oriented.

\begin{figure*}
\centering
\includegraphics[width = 0.9\textwidth]{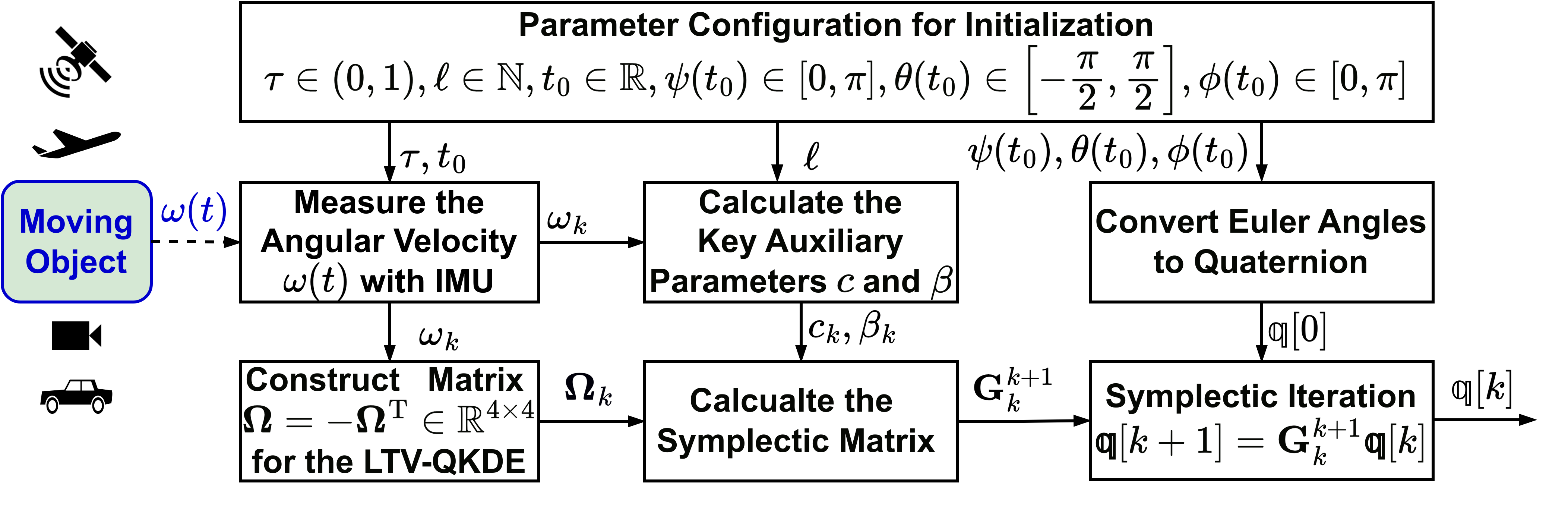} 
\caption{Flow chart of the EoESGA for solving the LTV-QKDE with accuracy parameter $\ell$.}
\label{fig-EoESGA-Solving-System}
\end{figure*}

The main purpose of this paper is to propose \textit{even order explicit symplectic geometric algorithms} (EoESGA) for solving the QKDE with Pad\'{e} approximation and Cayley transform based on the theorems proved for the purpose of engineering applications instead of pure theoretic exploration. The flow chart of solving the LTV-QKDE with the EoESGA via Pad\'{e} approximation and mirrored Cayley transform is given in \Fig \ref{fig-EoESGA-Solving-System}. There are six key steps for solving the LTV-QKDE with the EoESGA parameterized by the integer $\ell$, which will be discussed in the following sections.

The contents of this paper are organized as follows:
The preliminaries are given in Section \ref{sec-Preli}. Section \ref{sec-SDS4QKDE} deals with the SDS for the QKDE. In Section \ref{sec-EoESGA} we design the SGA and analyze the computational complexity. Section \ref{sec-ValiVeri} deals with performance evaluations. Finally, Section \ref{sec-Conclusion} presents the conclusions.

In order to help the reader to recognize the abbreviations, the nomenclatures are summarized in \Tab \ref{Tab-nomenclature}.
\begin{table}[h]
\centering
\caption{Nomenclatures} \label{Tab-nomenclature}
\begin{tabular}{lp{6cm}}
\hline
\textbf{Abbreviation}  & \textbf{Interpretation} \\
\hline 
QKDE &  quaternion kinematical differential equation \\
IMU  & inertial measure unit \\
A-QKDE & autonomous QKDE, or LTI-QKDE \\
NA-QKD & non-autonomous QKDE, or LTV-QKDE\\
LTI-QKDE & linear time-invariant QKDE \\
LTV-QKDE & linear time-varying QKDE \\
TFDS     & traditional finite difference scheme \\
SDS & symplectic difference scheme  \\
AS  & analytical solution \\
NS  & numerical solution \\
SGA & symplectic geometric algorithm\\
ESGA & explicit symplectic geometric algorithm\\
EoESGA & even order ESGA \\
DPA  & diagonal Pad\'{e} approximation \\
TCVC & time complexity vector of computation \\
\hline
\end{tabular}
\end{table}

\section{Preliminaries} \label{sec-Preli}

\subsection{Symplectic Transition Matrix}

Let $\tau$ be the time step and $\vec{z}[k] = \vec{z}(t_0+k\tau)$ be the sample value at the discrete time $t_k = t_0+k \tau$. 
The SDS for the Hamiltonian canonical equation, see Appendix \ref{appendix-Hamilton} and (\ref{eq-Heq}), is given by
\begin{equation}
\left\{
\begin{aligned}
&\vec{z}[k+1] = \mathcal{G}_\tau (\vec{z}[k]), \quad k = 0, 1, 2, \cdots 
\\
&\vec{z}[0] = \trsp{[p_1(t_0), \cdots, p_N(t_0), q_1(t_0), \cdots, q_N(t_0)]}
\end{aligned}
\right.
\end{equation}
where $\vec{z}[0]$ is the initial condition, $ \mathcal{G}_\tau:   \ES{R}{2N}{1}\to \ES{R}{2N}{1}$ 
is the \textit{transition operator}, also named as the \textit{transition mapping}, and
\begin{equation} \label{eq-G-trans-map}
 \mat{G}^{k+1}_{k}(\tau) = \fracpde{\vec{z}[k+1]}{\vec{z}[k]} = \fracpde{(\mathcal{G}_\tau \vec{z}[k])}{\vec{z}[k]}\in \ES{R}{2N}{2N},
 \end{equation}
is the \textit{transition matrix} (Jacobi matrix) of $\mathcal{G}_\tau$ 
at $\vec{z}[k]$  
   such that $\mat{G}^{k+1}_{k}(\tau)$  is symplectic, viz.,
\begin{equation}\label{eq-SpMat-def}
\trsp{\mat{G}^{k+1}_{k}(\tau)} \cdot \mat{J}_{2N} \cdot \mat{G}^{k+1}_{k}(\tau) = \mat{J}_{2N}.
\end{equation}
The key problem of designing SGA is to determine the $\mathcal{G}_\tau$ or the corresponding  Jacobi matrix $\mat{G}^{k+1}_{k}(\tau)$ defined by \eqref{eq-G-trans-map}. In this paper, we  have $\mathcal{G}_\tau  = \mat{G}^{k+1}_{k}(\tau)$ since (\ref{eq-QKDE}) is a linear ODE. 
For the SDS of LTI-QKDE, the notation $\mat{G}^{k+1}_{k}(\tau)$ will be replaced by  $\mat{G}(\tau)$ since it does not depend on the time index $k$. In this paper, we will also take the notations $\mat{G}^{k+1}_{k}(\ell,\tau)$ and $\mat{G}(\ell, \tau)$ where $\ell$ is the order parameter for accuracy.

\subsection{Pad\'{e} Approximation and Symplectic Difference Scheme }

For non-negative integers $\ell$ and $m$, the $(\ell+m)$-th order Pad\'{e} approximation to $\exp(x)$ is defined by \cite{Baker1996PadeApprox,Cuyt1984Pade,Feng2010}
\begin{equation} \label{eq-exp-approx-Pade}
\exp(x) =  \frac{N_{\ell m}(x)}{D_{\ell m}(x)}  + \mathcal{O}(\abs{x}^{\ell +m+1})
\end{equation}
where
\begin{equation} \label{eq-N-D-Pade}
\left\{
\begin{aligned}
N_{\ell m}(x) &= \sum^m_{k=0} \frac{(\ell +m -k)! m!}{(\ell +m)! k!(m-k)!}x^k, \\
D_{\ell m}(x) &= \sum^m_{k=0} \frac{(\ell +m -k)! \ell!}{(\ell +m)! k!(\ell-k)!}(-x)^k.
\end{aligned}
\right.
\end{equation}
The \textit{diagonal Pad\'{e} approximation} (DPA) is specified by the condition $\ell = m$, namely
\begin{equation} \label{eq-exp-Pade-diagonal}
\exp(x) =  \frac{N_{\ell \ell}(x)}{D_{\ell \ell}(x)}  + \mathcal{O}(\abs{x}^{2\ell+1})
\end{equation}
where
\begin{equation} \label{eq-N-D-diagonal}
\left\{
\begin{aligned}
N_{\ell\ell}(x) &= \sum^\ell_{k=0} \frac{(2\ell -k)! \ell!}{(2\ell)! k!(\ell-k)!}x^k \\
 D_{\ell\ell}(x) &= \sum^\ell_{k=0} \frac{(2\ell -k)! \ell!}{(2\ell)! k!(\ell-k)!}(-x)^k .
\end{aligned}
\right.
\end{equation}
Let
\begin{equation} \label{eq-P-ell-x-def}
P_{\ell}(x) 
=\sum^\ell_{k=0} \dfrac{(2\ell -k)!}{(2\ell)!}\cdot\dfrac{\ell!}{(\ell-k)!}\cdot\dfrac{x^k}{k!},
\end{equation}
then both the nominator $N_{\ell\ell}(x)$ and denominator $D_{\ell\ell}(x)$ can be expressed by the $P_\ell(x)$, namely,
\begin{equation}
N_{\ell\ell}(x)= P_\ell(x), D_{\ell\ell}(x) = P_\ell(-x).
\end{equation}
Simple algebraic operations on factorials show that
\begin{equation} \label{eq-P-ell-x}
\begin{split}
P_{\ell}(x) 
&= 1 + \sum^{\ell}_{k=1}\frac{(2\ell-k)!}{(2\ell)!}\cdot\frac{\ell!}{(\ell-k)!}\cdot\frac{x^k}{k!} \\
&= 1+ \sum^{\ell-1}_{k=0}x^{k+1}\prod^{k}_{r=0} \frac{\ell-r}{(2\ell-r)(r+1)} 
\end{split}
\end{equation}
It should be pointed out that the factorials in \eqref{eq-P-ell-x-def} for $k=0$ has been removed in \eqref{eq-P-ell-x} in order to avoid the  overflow problem in numerical computations when $2\ell > 10$. 
Put 
\begin{equation} \label{eq-eta-l-r}
\eta^\ell_r = \frac{\ell-r}{(2\ell-r)(r+1)}, \quad r\in \set{0,1,\cdots,\ell-1},
\end{equation}
then we can obtain
\begin{equation} \label{eq-P-ell-eta-x}
P_{\ell}(x) 
= 1+ \sum^{\ell-1}_{k=0}x^{k+1}\prod^k_{r=0}\eta^\ell_r
\end{equation}
Given the parameter $\ell$, \eqref{eq-P-ell-eta-x} can be used to generate the expression of $P_\ell(x)$. 
For illustration, we have
\begin{equation}
\left\{
\begin{aligned}
P_1(x)  &=   1+\frac{x}{2} \\
P_2(x)  &=   1 + \frac{x}{2} + \frac{x^2}{12} \\
P_3(x)  &=   1+ \frac{x}{2} + \frac{x^2}{10} + \frac{x^3}{120}\\
P_4(x)  &=   1+ \frac{x}{2} + \frac{3x^2}{28} + \frac{x^3}{84} + \frac{x^4}{1680}\\
P_5(x) &= 1 + \frac{x}{2} + \frac{x^2}{9} + \frac{x^3}{72} + \frac{x^4}{1008} + \frac{x^5}{30240}\\
P_6(x) &= 1+\frac{x}{2} + \frac{5x^2}{44} + \frac{x^3}{66} +
\frac{x^4}{792}  +  \frac{x^5}{15840}  + \frac{x^6}{665280}
\end{aligned} 
\right.
\end{equation}
with the help of \eqref{eq-eta-l-r} and \eqref{eq-P-ell-eta-x}.

Let
\begin{equation} \label{eq-g-l-x-def}
\scrd{g}{DPA}^\ell(x) = \frac{N_{\ell\ell}(x)}{D_{\ell\ell}(x)}=\frac{P_{\ell}(x)}{P_{\ell}(-x)}
\end{equation}
then the DPA can be written by 
\begin{equation} \label{eq-g-l-x}
\exp(x)\sim \scrd{g}{DPA}^\ell(x) + \mathcal{O}(\abs{x}^{2\ell+1})
\end{equation}
according to \eqref{eq-exp-Pade-diagonal} and \eqref{eq-N-D-diagonal}.
The function $\scrd{g}{DPA}^\ell(x)$ for approximating the $\exp(x)$ can be used to construct the SDS of interest. Actually, we have the following theorem \cite{Feng2010}.
\begin{thm} \label{SDS-2L}
For the linear Hamiltonian system
\begin{equation} \label{eq-LinearHS}
\fracode{\vec{z}}{t} = \mat{F} \vec{z}, \quad \mat{F} = \inv{\mat{J}} \mat{C} , \quad \trsp{\mat{C}} = \mat{C},
\end{equation}
where $\mat{F} = \inv{\mat{J}} \mat{C}$ is infinitesimal symplectic, i.e.,
\begin{equation}
\mat{J}\mat{F} + \trsp{\mat{F}}\mat{J} = \mat{O}.
\end{equation}
Let
\begin{equation}
\begin{split}
\scrd{g}{DPA}^\ell(\tau\mat{F}) 
&= P_\ell(\tau \mat{F}) \inv{[P_\ell(-\tau \mat{F})]}
\end{split}
\end{equation}
by replacing the symbol $x$ in \eqref{eq-g-l-x} with the matrix $\tau\mat{F}$, 
then the difference schemes 
\begin{equation}
\vec{z}[k+1] = \scrd{g}{DPA}^{\ell}(\tau \mat{F})\vec{z}[k], \quad \ell = 1, 2, \cdots 
\end{equation}
are symplectic of $2\ell$-th order accuracy for step size $\tau$.
\end{thm}

\subsection{Cayley Transform and Euler's Formula}

We now cite two lemmas about mirrored Cayley transform and Euler's formula proved in \cite{Zhang2018QKDE}.  
\begin{lem}[\textbf{Generalized Euler's formula}]  \label{lem-Euler}
Suppose matrix $\mat{M}\in \ES{R}{n}{n}$ is skew-symmetric, i.e., $\trsp{\mat{M}} = -\mat{M}$, and there exists
a positive constant $\gamma$ such that $\mat{M}^2 = -\gamma^2 \mat{I}$. Let $\hat{\mat{M}} = {\gamma}^{-1}\mat{M}$, then $\hat{\mat{M}}^2 = -\mat{I}$. 
 For any $y\in \mathbb{R}$, we have
\begin{equation}
\me^{y\mat{M}} = \cos(y\gamma)\mat{I}
+ \sin(y\gamma)\hat{\mat{M}}.
\end{equation}
\end{lem}

\begin{lem}[\textbf{Cayley-Euler formula}]  \label{lem-Cayley-Euler}
Suppose matrix $\mat{M}\in \ES{R}{n}{n}$ is skew-symmetric, i.e., $\trsp{\mat{M}} = -\mat{M}$, and there exists
a positive constant $\gamma$ such that $\mat{M}^2 = -\gamma^2 \mat{I}$. Let $\hat{\mat{M}} = {\gamma}^{-1}\mat{M}$, then $\hat{\mat{M}}^2 = -\mat{I}$. 
Let 
\begin{equation} \label{eq-Cayley-trans-def}
\scrd{\varphi}{mct}(\lambda) = \frac{1+\lambda}{1-\lambda}=(1+\lambda)\inv{(1-\lambda)}
\end{equation} be the
\textit{mirrored Cayley transform}, then for any $y\in \mathbb{R}$ the Cayley-Euler formula
\begin{equation}
\begin{split}
\scrd{\varphi}{mct}(y\mat{M})
&= \frac{\mat{I}+y\mat{M} }{\mat{I}-y\mat{M}}
=
\frac{1}{1+\alpha}[(1-\alpha)\mat{I} +2y\mat{M}]\\
&=\cos \delta(y, \gamma) \mat{I} + \sin \delta(y, \gamma) \hat{\mat{M}}\\
&= \me^{\delta(y, \gamma)\mat{M}}
\end{split}
\end{equation}
holds, in which 
\begin{equation}
\begin{cases}
\delta = \delta(y, \gamma) = 2\cdot \arctan (y\gamma)\\
\alpha =\alpha(y,\gamma)= y^2\gamma^2
\end{cases}
\end{equation}
Furthermore, $\scrd{\varphi}{mct}(y\mat{M})$ is an orthogonal matrix.
\end{lem}

For more information about the mirrored Cayley transform and its properties, please see the Appendix \ref{sec-appen-cayley}.

\section{Symplectic Difference Schemes for QKDE}\label{sec-SDS4QKDE}

\subsection{Relation of Pad\'{e} Approximation and Cayley Transform }

Generally, the Pad\'{e} approximation involves the ratio of  polynomials with the same order. It is interesting that the expression of Pad\'{e} approximation can be simplified into the mirrored Cayley transform for solving the QKDE. 
We now prove some interesting lemmas and theorems for constructing the SDS for LTI-QKDE.

\begin{lem}[\textbf{Pad\'{e}-Cayley}]  \label{lem-PadeSpMat}
If there exists some positive constant $c$ such that $x^2 = -c$, then each diagonal Pad\'{e} approximation $\scrd{g}{DPA}^\ell(x)$ can be represented by the  mirrored Cayley transform
\begin{equation} \label{eq-g-ell-x-ratio}
\scrd{g}{DPA}^\ell(x) = \frac{1 + \beta(\ell,c)x}{1-\beta(\ell,c)x} = \scrd{\varphi}{mct}\left(\beta(\ell,c) x\right),
\end{equation}
where $\beta(\ell,c)$
is a constant determined by $\ell$ and $c$.
\end{lem}

\noindent\textsc{Proof}: Let 
$\mfloor{x} = m$ in which $\exists m \in \mathbb{Z}$ such that $m\le x < m +1$, and 
\begin{equation} \label{eq-def-s1-s2}
s_1 = \mfloor{(\ell-1)/2}, \quad s_2 = \mfloor{\ell/2},
\end{equation}
be two integer parameters, we immediately have
\begin{equation} \label{eq-s1-s2-relation}
s_2 
= s_1 + 1 - \mod(\ell,2) =
\begin{cases}
s_1, & \quad 2\nmid \ell;\\
s_1 + 1, & \quad 2\mid \ell. 
\end{cases}
\end{equation}
When $x^2 = -c$, equation \eqref{eq-P-ell-x} shows that
\begin{equation} \label{eq-P-ell-lin}
\begin{split}
P_\ell(x) 
&=  1 + \sum^{\ell-1}_{k=0} x^{k+1} \prod^k_{r=0}\eta^\ell_r\\
&= 1 + \sum^{s_1}_{j=0} x^{2j+1} \prod^{2j}_{r=0}\eta^\ell_r + \sum^{s_2}_{j=1} x^{2j} \prod^{2j-1}_{r=0}\eta^\ell_r\\
&=1  + \sum^{s_2}_{j=1} (-c)^j \prod^{2j-1}_{r=0}\eta^\ell_r
 + x\sum^{s_1}_{j=0} (-c)^j \prod^{2j}_{r=0}\eta^\ell_r.
\end{split}
\end{equation} 
Let
\begin{equation}\label{eg-ab-lj}
\left\{
\begin{split}
a^\ell_j &= \prod^{2j}_{r=0}\eta^\ell_r, \quad 0 \le j \le s_1   \\
b^\ell_j &= \prod^{2j-1}_{r=0}\eta^\ell_r,  \quad 1 \le j \le s_2, \\
b^\ell_0 & = 1,
\end{split}
\right.
\end{equation}
and
\begin{equation} \label{eq-n-d-x}
\left\{
\begin{aligned}
n(s_1,x) &= \sum^{s_1}_{j=0} a^\ell_j x^j\\ 
d(s_2,x) &= \sum^{s_2}_{j=0} b^\ell_j x^j
\end{aligned} 
\right.
\end{equation}
By substituting \eqref{eg-ab-lj} and \eqref{eq-n-d-x} into \eqref{eq-P-ell-lin}, we have
\begin{equation} \label{eq-P-l-x}
\begin{split}
P_\ell(x) 
&= \sum^{s_2}_{j=0} b^\ell_j \cdot (-c)^j +x \sum^{s_1}_{j=1} a^\ell_j \cdot (-c)^j\\
&=  d(s_2,-c) + x \cdot n(s_1,-c).
\end{split}
\end{equation}
Obviously,  \eqref{eq-P-l-x} is the simplified version of \eqref{eq-P-ell-x} under the condition $x^2=-c$. Let 
\begin{equation} \label{eq-beta-l-c-def}
\beta(\ell,c) = \dfrac{n\left(s_1,-c\right)}{d\left(s_2,-c\right)}
=\dfrac{\sum\limits^{s_1}_{j=0} a^\ell_j (-c)^j }{ \sum\limits^{s_2}_{j=0} b^\ell_j (-c)^j},
\end{equation}
by substituting \eqref{eq-P-l-x} and \eqref{eq-beta-l-c-def} into
\eqref{eq-g-l-x-def}, we have 
\begin{equation} \label{eq-ratio}
\begin{aligned}
\scrd{g}{DPA}^\ell(x)&=\frac{P_\ell(x)}{P_\ell(-x)} = \frac{d(s_2,-c) + x \cdot n(s_1,-c) }{d(s_2,-c) - x \cdot n(s_1,-c)}\\
&= \frac{1 + \beta(\ell,c)x}{1-\beta(\ell,c)x} 
= \scrd{\varphi}{mct}\left(\beta(\ell,c) x\right). 
\end{aligned}
\end{equation}
according to \eqref{eq-Cayley-trans-def}. Therefore, we can deduce that \eqref{eq-g-ell-x-ratio} holds by \eqref{eq-g-l-x} and \eqref{eq-ratio}.  Q.E.D.

\subsection{Procedure for Computing $\displaystyle \beta(\ell,c)$}

\begin{figure*}[htpb]
\centering
\includegraphics[width = 0.80\textwidth]{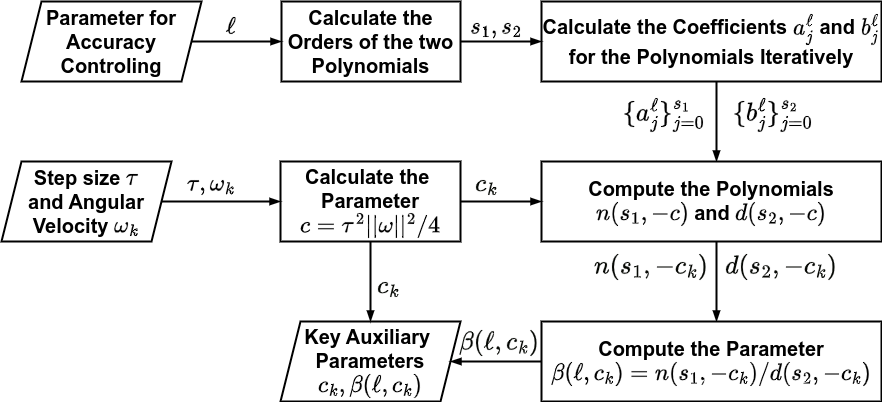} 
\caption{Flow chart of computing the key auxiliary parameters $c$ and $\beta$.}
\label{fig-CalcBeta}
\end{figure*}

The flow chart calculating the auxilary parameters $c$ and $\beta$ is illustrated in \Fig \ref{fig-CalcBeta}.
There are two key tasks for computing $\beta(\ell,c)$:  
\begin{itemize}
\item determine the coefficients $a^\ell_j$ and $b^\ell_j$;
\item calculate the values of polynomials $n(s_1,-c)$ and $d(s_2, -c)$ when $c$ is determined. 
\end{itemize}
For the coefficients of interest, we can use iterative algorithms to generate them. For the computations of polynomials, we can use the famous rule invented by  Jiushao Qin and Horne independently \cite{Jiushao1247, Horner, KnuthTAOCP2}, see \Algr \ref{alg-Polynomial} in Appendix \ref{appendix-poly}. 

There are two typical ways to compute the coefficients $a^\ell_j$ and $b^\ell_j$ for the polynomials $n(s_1,-c)$ and $d(s_2,-c)$.
\begin{itemize}
\item[1)] \textbf{Parallel Iterative method}:
According to the definition of $a^\ell_j$, we can find the following iterative formulas for the coefficients $a^\ell_j$: 
\begin{equation} \label{eq-i-a}
\left\{
\begin{split}
&a^\ell_{j+1} 
= a^\ell_{j}\cdot\eta^\ell_{2j+1}\cdot
\eta^\ell_{2j+2}, \quad 0 \le j \le s_1 -1; \\
&a^\ell_0  = \frac{1}{2}.
\end{split}
\right.
\end{equation}
Similarly, we can obtain
\begin{equation} \label{eq-i-b}
\left\{
\begin{aligned}
&b^\ell_{j+1} 
= b^\ell_{j} \cdot \eta^\ell_{2j}\cdot \eta^\ell_{2j+1}, \quad 0\le  j \le s_2 -1 \\
&b^\ell_0 = 1. 
\end{aligned}
\right.
\end{equation}

\item[2)] \textbf{Alternative Iterative Method}:
We can also obtain the following alternative iterative formula for the coefficients $a^\ell_j$ and $b^\ell_j$
\begin{equation} \label{eq-ai-ab}
\begin{cases}
b^\ell_0 = 1,  &\\
a^\ell_0 = \frac{1}{2}, &\\
b^\ell_{j+1} =a^\ell_j \cdot \eta^\ell_{2j+1}, & 0 \le j \le s_1 -1\\
a^\ell_{j+1} =b^\ell_{j+1}\cdot \eta^\ell_{2j+2},& 0 \le j \le s_1 -1\\
b^\ell_{s_2}= a^\ell_{s_1}\cdot \eta^\ell_{2s_1+1}, &  \hbox{for even $\ell$ }.  
\end{cases}
\end{equation}
\end{itemize}
It is easy to find that the computations in alternative iterative formula (\ref{eq-ai-ab}) are more efficient than those in (\ref{eq-i-a}) and (\ref{eq-i-b}) since only $\eta^\ell_{2j+1}$ and $\eta^\ell_{2j+2}$ will be computed. 

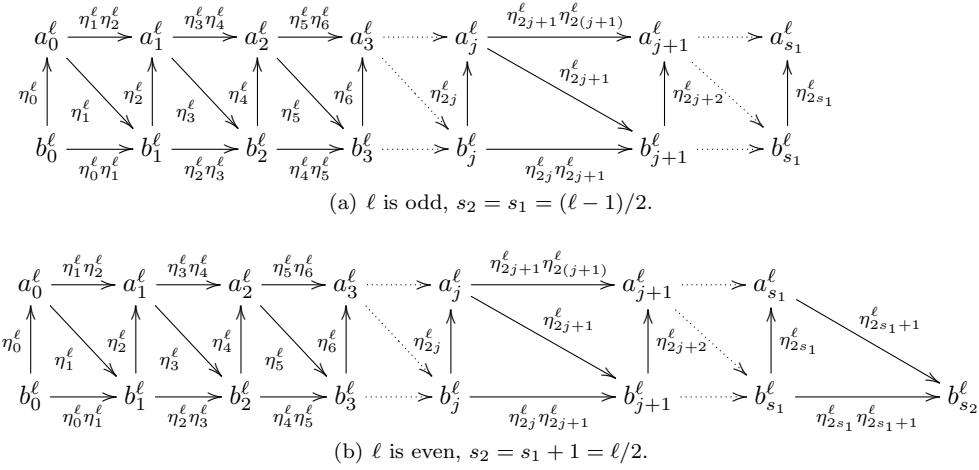
\begin{figure*}[htbp]
\begin{center}
\subfigure[$\ell$ is odd, $s_2 = s_1=(\ell-1)/2$.]{
        $$
\xymatrix{
a^\ell_0\ar[r]^{\eta^\ell_1\eta^\ell_2}\ar[dr]_{\eta^\ell_1} &
a^\ell_1\ar[r]^{\eta^\ell_3\eta^\ell_4}\ar[dr]_{\eta^\ell_3} & 
a^\ell_2\ar[r]^{\eta^\ell_5\eta^\ell_6}\ar[dr]_{\eta^\ell_5} &
a^\ell_3
\ar@{.>}[r]^{ }\ar@{.>}[dr]|-{} & 
a^\ell_j\ar[rr]^{\eta^\ell_{2j+1}\eta^\ell_{2(j+1)}}\ar[drr]^{\eta^\ell_{2j+1}} &
&
a^\ell_{j+1}\ar@{.>}[r]^{}\ar@{.>}[dr]_{}&
a^\ell_{s_1}
&
&\\
b^\ell_0\ar[r]_{\eta^\ell_0\eta^\ell_1}\ar[u]^{\eta^\ell_0} &
b^\ell_1\ar[r]_{\eta^\ell_2\eta^\ell_3}\ar[u]^{\eta^\ell_2} &
b^\ell_2\ar[r]_{\eta^\ell_4\eta^\ell_5}\ar[u]^{\eta^\ell_4} &
b^\ell_3\ar@{.>}[r]_{ }\ar[u]^{\eta^\ell_6} & 
b^\ell_j\ar[rr]_{\eta^\ell_{2j}\eta^\ell_{2j+1}}\ar[u]^{\eta^\ell_{2j}} &
&
b^\ell_{j+1}\ar@{.>}[r]\ar[u]_{\eta^\ell_{2j+2}}&
b^\ell_{s_1}\ar[u]_{\eta^\ell_{2s_{1}}}
&
&\\
}
$$
        \label{fig-odd-even}
}
\subfigure[$\ell$ is even, $s_2 = s_1 + 1=\ell/2$.]{
 $$
\xymatrix{
a^\ell_0\ar[r]^{\eta^\ell_1\eta^\ell_2}\ar[dr]_{\eta^\ell_1} &
a^\ell_1\ar[r]^{\eta^\ell_3\eta^\ell_4}\ar[dr]_{\eta^\ell_3} & 
a^\ell_2\ar[r]^{\eta^\ell_5\eta^\ell_6}\ar[dr]_{\eta^\ell_5} &
a^\ell_3
\ar@{.>}[r]^{ }\ar@{.>}[dr]|-{} & 
a^\ell_j\ar[rr]^{\eta^\ell_{2j+1}\eta^\ell_{2(j+1)}}\ar[drr]^{\eta^\ell_{2j+1}} &
&
a^\ell_{j+1}\ar@{.>}[r]^{}\ar@{.>}[dr]_{}&
a^\ell_{s_1}\ar[drr]^{\eta^\ell_{2s_{1}+1}}
&
&\\
b^\ell_0\ar[r]_{\eta^\ell_0\eta^\ell_1}\ar[u]^{\eta^\ell_0} &
b^\ell_1\ar[r]_{\eta^\ell_2\eta^\ell_3}\ar[u]^{\eta^\ell_2} &
b^\ell_2\ar[r]_{\eta^\ell_4\eta^\ell_5}\ar[u]^{\eta^\ell_4} &
b^\ell_3\ar@{.>}[r]_{ }\ar[u]^{\eta^\ell_6} & 
b^\ell_j\ar[rr]_{\eta^\ell_{2j}\eta^\ell_{2j+1}}\ar[u]^{\eta^\ell_{2j}} &
&
b^\ell_{j+1}\ar@{.>}[r]\ar[u]_{\eta^\ell_{2j+2}}&
b^\ell_{s_1}\ar[rr]_{\eta^\ell_{2s_{1}}\eta^\ell_{2s_{1}+1}}\ar[u]_{\eta^\ell_{2s_{1}}}
&
&b^\ell_{s_2}\\
}
$$
        \label{fig-ell-even}
}
\end{center}
\caption{Iterative processes for calculating the coefficients $a^\ell_j$ and $b^\ell_j$ correspond to paths on graph.}
\label{fig-iter-graph}
\end{figure*}

The iterative algorithms for generating $a^\ell_j$ and $b^\ell_j$ are shown in Figure \ref{fig-iter-graph}. We can find some interesting facts from this figure:
\begin{itemize}
\item The structures of the graph for  odd $\ell$ and even $\ell$ are different. If $\ell$ is even, then $s_2 = s_1 + 1 = \ell/2$ by \eqref{eq-s1-s2-relation} and we need an extra iterative step from $a^\ell_{s_1}$ to $b^\ell_{s_2}$ when compared with the case of odd $\ell$. 
\item For the parallel iterative method, the paths for computing $\set{a^\ell_j}^{s_1}_{j=0}$ and $\set{b^\ell_j}^{s_2}_{j=0}$ are  $a^\ell_0 \rightarrow a^\ell_1 \rightarrow  \cdots \rightarrow a^\ell_j \rightarrow \cdots \rightarrow a^\ell_{s_1}$ and  $b^\ell_0 \rightarrow b^\ell_1 \rightarrow  \cdots \rightarrow b^\ell_j \rightarrow \cdots \rightarrow b^\ell_{s_2}$ respectively. It is obvious that these two paths are independent and the computations involves lots of the products $\eta^\ell_{2j+1}\eta^\ell_{2j+2}$ and $\eta^\ell_{2j}\eta^\ell_{2j+1}$. 
\item 
 For the alternative iterative method, the path for computing the coefficients of interest is $b^\ell_0 \rightarrow a^\ell_0 \rightarrow b^\ell_1 \rightarrow a^\ell_1 \rightarrow \cdots \rightarrow b^\ell_j \rightarrow a^\ell_j \rightarrow \cdots \rightarrow b^\ell_{s_1} \rightarrow a^\ell_{s_1} \rightarrow b^\ell_{s_2}$ when $\ell$ is even or $b^\ell_0 \rightarrow a^\ell_0 \rightarrow b^\ell_1 \rightarrow a^\ell_1 \rightarrow \cdots \rightarrow b^\ell_j \rightarrow a^\ell_j \rightarrow \cdots \rightarrow b^\ell_{s_1} \rightarrow a^\ell_{s_1}$ when $\ell$ is odd.
\end{itemize}

With the help of $a^\ell_j$ and $b^\ell_j$, we can find the expression of $\beta(\ell,c)$ with fixed $\ell$. For $1\le \ell \le 6$, the $\beta(\ell,c)$ are illustrated in \Tab \ref{Tab-beta-Taylor}.  
Obviously, $\beta(1,c)\equiv 1/2$ does not depend on the constant $c$. The series expansion  of $\beta(\ell,c)$ at $c=0$ are also demonstrated in \Tab \ref{Tab-beta-Taylor}.
\begin{table*}[htbp]
\centering
\caption{Expressions of $\beta(\ell,c)$ for $1\le \ell\le 6$}
\label{Tab-beta-Taylor}
\begin{spacing}{2.2}
\resizebox{\textwidth}{!}{
\begin{tabular}{clll}
\hline 
$\ell$ & $\beta(\ell,c)=\dfrac{n(s_1,-c)}{d(s_2,-c)}$  & 
Taylor series of $\beta(\ell,c)$ at $c = 0$ & DOC  \\
\hline \hline
$1$ & $ 
\dfrac{1}{2}$    & $\dfrac{1}{2}$  & $[0, +\infty)$\\
$2$ & $ \dfrac{\dfrac{1}{2}}{1-\dfrac{1}{12}c}$    & $ \dfrac{1}{2} +  \dfrac{c}{24} + \dfrac{c^2}{288} + \dfrac{c^3}{3456} + \dfrac{c^4}{41472} + \dfrac{c^5}{497664} + \dfrac{c^6}{5971968} + \mathcal{O}(c^7)$  & $[0, 12)$\\
$3$ & $ \dfrac{\dfrac{1}{2}-\dfrac{1}{120}c}{1-\dfrac{1}{10}c}$    & $\dfrac{1}{2}+\dfrac{c}{24}+\dfrac{c^2}{240}+\dfrac{c^3}{2400}+  \dfrac{c^4}{24000}
+   \dfrac{c^5}{240000} + \dfrac{c^6}{2400000}  +  \mathcal{O}\left(c^7\right)$ & $[0,10)$\\
$4$ & $ \dfrac{\dfrac{1}{2}-\dfrac{1}{84}c}{1-\dfrac{3}{28}c+\dfrac{1}{1680}c^2}$    & $\dfrac{1}{2}+\dfrac{c}{24}+\dfrac{c^2}{240}+\dfrac{17 c^3}{40320}+\dfrac{241 c^4}{5644800}+ \dfrac{41c^5}{9483264} 
+ \dfrac{29063c^6}{66382848000} +\mathcal{O}\left(c^7\right)$ & $\left[0, \dfrac{12650}{1281}\right)$\\
$5$ & $ \dfrac{\dfrac{1}{2}-\dfrac{1}{72}c+\dfrac
{1}{30240}c^2}{1 -\dfrac{1}{9}c + \dfrac{1}{1008} c^2}$    & $\dfrac{1}{2}+\dfrac{c}{24}+\dfrac{c^2}{240}+\dfrac{17 c^3}{40320}+\dfrac{31 c^4}{725760}+\dfrac{1583 c^5}{365783040}+ \dfrac{2887c^6}{6584094720} + 
\mathcal{O}\left(c^7\right)$ & $\left[0, \dfrac{2349}{238}\right)$\\
$6$& $ \dfrac{\dfrac{1}{2}-\dfrac{1}{66}c + \dfrac{1}{15840}c^2}{1-\dfrac{5}{44}c+\dfrac{1}{792}c^2-\dfrac{1}{665280}c^3}$    & $\dfrac{1}{2}+\dfrac{c}{24}+\dfrac{c^2}{240}+\dfrac{17 c^3}{40320}+\dfrac{31 c^4}{725760}+\dfrac{691 c^5}{159667200}+\dfrac{388151 c^6}{885194956800}+\mathcal{O}\left(c^7\right)$ & $\left[0, \dfrac{10294}{1043}\right)$\\
\hline
\end{tabular}
}
\end{spacing}
\end{table*}

\subsection{Impacts of parameters $\ell$ and $c$ on $\beta(\ell,c)$ }
It is obvious that the transition matrix $\mat{G}^{k+1}_k(\ell,\tau)$ depends on the order parameter $\ell$ by the value of $\beta(\ell,c)$ essentially. Some expressions of $\beta(\ell,c)$ for $1\le \ell \le 6$ are demonstrated in \Tab \ref{Tab-beta-Taylor}. Note that for different $\ell$, the function $\beta(\ell,c)$ will have different domain of convergence (DOC) for parameter $c$.

\begin{figure*}[h]
\subfigure[$\beta(\ell,c)$ varies with $\ell$ for fixed $c$]{
    \includegraphics[width=0.5\textwidth,height=5cm]{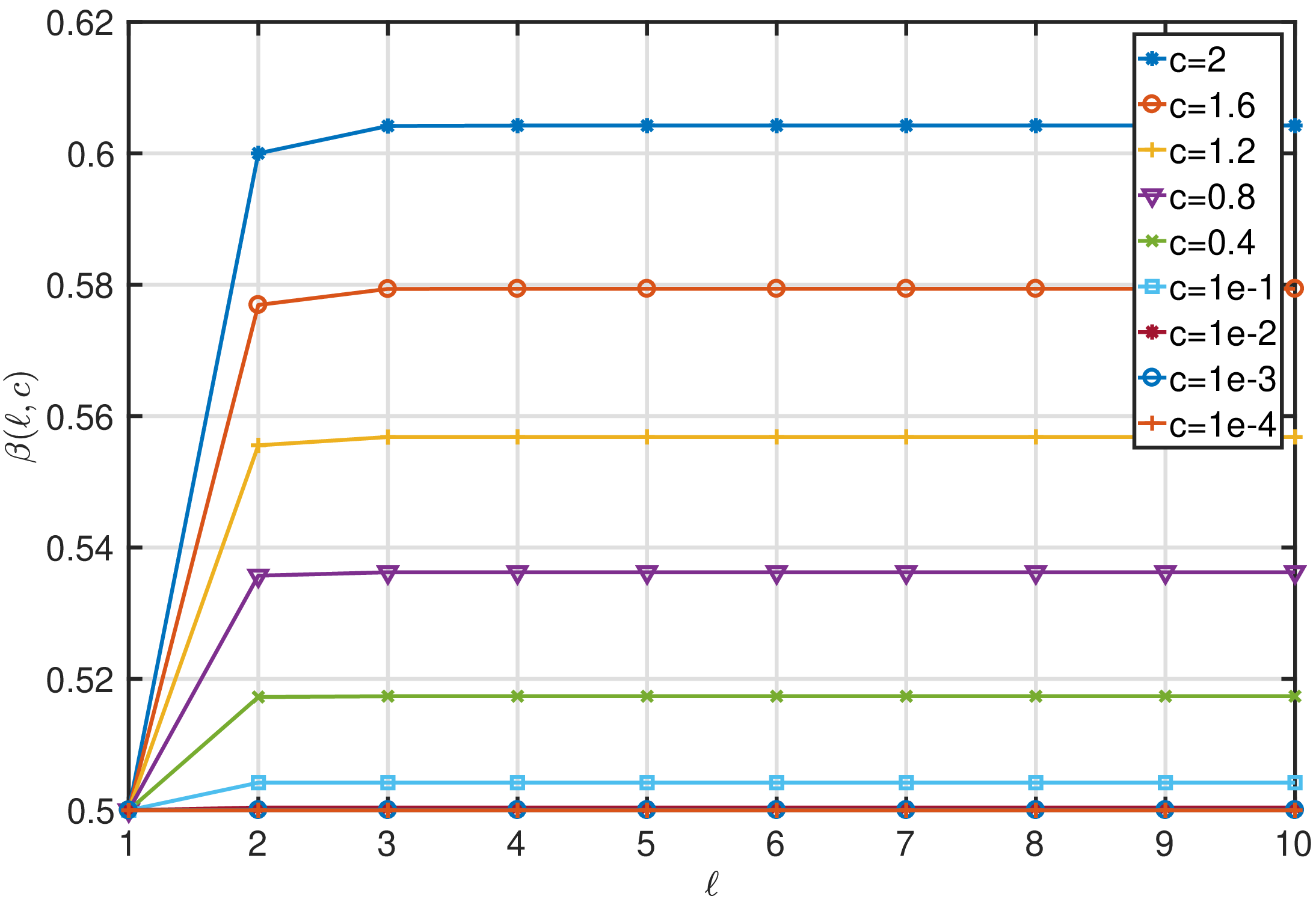}
    \label{fig-beta-ell}
}
\subfigure[$\beta(\ell,c)$ varies with $c$ for fixed $\ell$]{
    \includegraphics[width=0.5\textwidth,height=5cm]{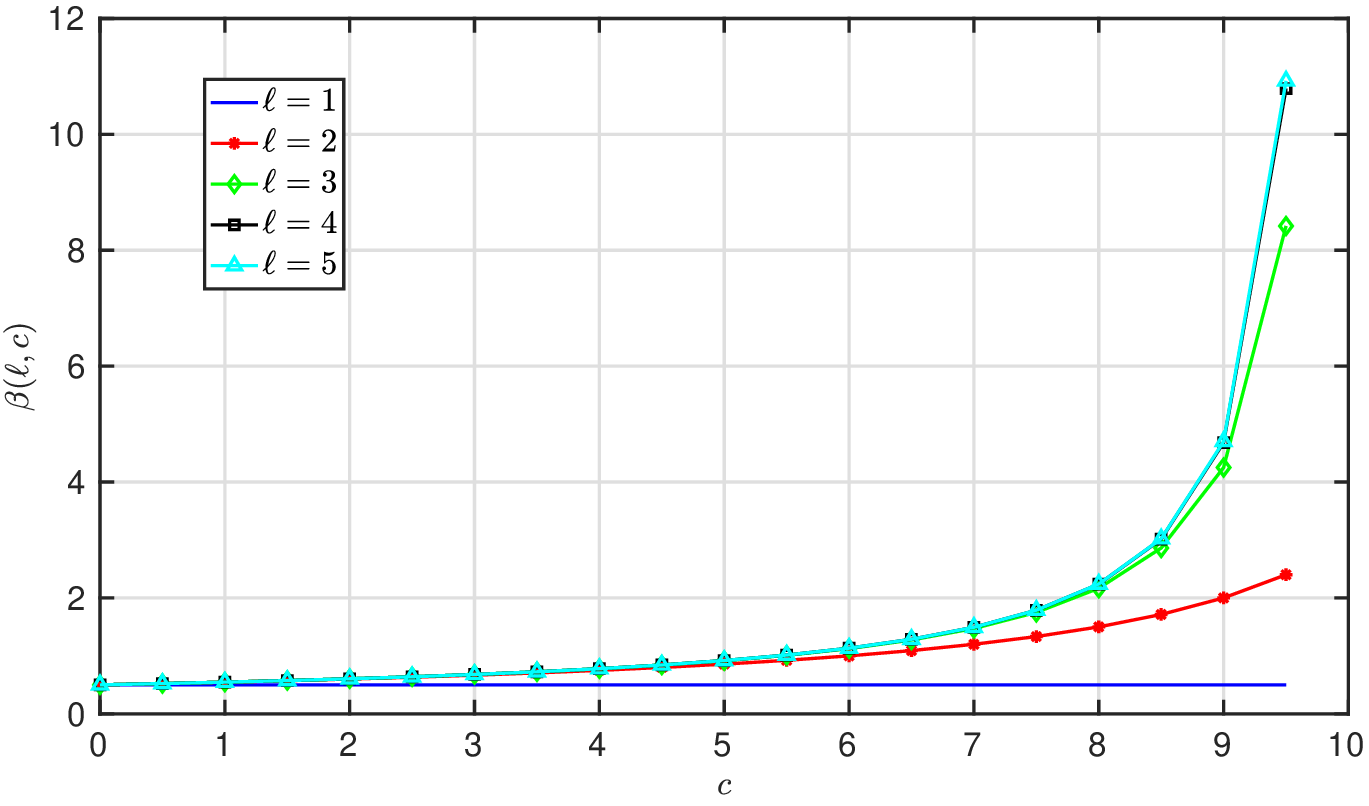}
    \label{fig-beta-c}
}
\caption{Impact of $\ell$ and $c$ on $\beta(\ell,c)$}
\label{fig-beta}
\end{figure*}

\Fig \ref{fig-beta} illustrates the $\beta(\ell,c)$ for different $\ell$ and $c$. \Fig \ref{fig-beta-ell} shows  that for $c\le 1.0\times 10^{-2}$ we have $\beta(\ell,c)\approx \beta(1,c) = \frac{1}{2}$. However, for $c\ge 0.1$, $\beta(\ell,c)$ varies with $c$ significantly. Moreover, for $\ell\ge 4$, $\beta(\ell,c)$ has  stable values for any $c$, which implies that $\ell \ge 4$ will be a satisfactory order parameter. Note that $\ell$ is a discrete variable and we use continuous lines for visualization in \Fig \ref{fig-beta-ell}.
On the other hand, \Fig \ref{fig-beta-c} shows that for $\beta(\ell,c)$ differs a lot when $c > 2.0$. Moreover, for $\ell \ge 4$, the $\beta(\ell,c)$ coincides well for any $c$. In general, it is a good choice for us to set $\ell = 4$ if we prefer high order algorithm.

\subsection{$2\ell$-th order SDS for LTI-QKDE}

\begin{lem} The diagonal Pad\'{e} approximation of order $\ell$ for the transition matrix  for the LTI-QKDE  is
\begin{equation} \label{eq-gMat}
\scrd{g}{DPA}^\ell\left(\frac{\tau}{2}\mat{\Omega}\right) 
=  \scrd{\varphi}{mct}\left( \beta(\ell,c)\frac{\tau}{2}  \mat{\Omega}\right) = \frac{\mat{I} + \beta(\ell,c)\frac{\tau}{2}  \mat{\Omega}}{\mat{I} - \beta(\ell,c)\frac{\tau}{2}  \mat{\Omega}}
\end{equation}
where
\begin{equation} \label{eq-c-def}
 c = \norm{\vec{\omega}}^2\tau^2/4.
\end{equation}
\end{lem}

\noindent \textsc{Proof}: Equation \eqref{eq-c-def} implies that  
$\left(\tau\mat{\Omega}/2\right)^2 =  -c \mat{I}$. Thus by Lemma \ref{lem-PadeSpMat} we can find that \eqref{eq-gMat} holds.

\begin{thm} For any time step $\tau$ and time-invariant vector $\vec{\omega}$, the transition matrix  for the LTI-QKDE is
\begin{equation}  \label{eq-mat-G-ell-tau}
\begin{split}
\mat{G}(\ell,\tau)
&= \scrd{g}{DPA}^\ell\left(\frac{\tau}{2}\mat{\Omega}  \right)\\
&= \frac{1}{1+\alpha(\ell,c)}  \left[ \left( 1 - \alpha(\ell,c)\right)\mat{I}
    +\tau \beta(\ell,c)\mat{\Omega}\right]\\
&= \cos\delta(\ell,\vec{\omega}, \tau)   \mat{I} + \sin \delta(\ell,\vec{\omega},\tau)  \hat{\mat{\Omega}},
\end{split}
\end{equation}
in which 
\begin{equation} \label{eq-parameters}
\left\{
\begin{aligned}
&\alpha(\ell, c) = c [\beta(\ell,c)]^2, \\
&\delta(\ell, \vec{\omega},\tau) = 2 \arctan (\beta(\ell,c)\norm{\vec{\omega}}\tau/2),     \\
&\hat{\mat{\Omega}} =\mat{\Omega}/\norm{\vec{\omega}}, \\
&\hat{\mat{\Omega}}^2 = -\mat{I}.
\end{aligned}
\right.
\end{equation}
\end{thm}

\noindent \textsc{Proof}: Let 
\begin{align*}
y = \tau\beta(\ell,c)/2, \quad
\gamma = \norm{\vec{\omega}}, \quad 
\delta  = 2 \arctan (y\gamma),
\end{align*}
then  $\mat{\Omega}^2 = -\gamma^2 \mat{I}$ and  
\begin{equation*}
\alpha(\ell,c) = y^2 \gamma^2 =  c [\beta(\ell,c)]^2.
\end{equation*} 
Let $\mat{M} = \mat{\Omega}$, then this theorem follows from the
Lemma \ref{lem-Cayley-Euler} directly. Q.E.D.

\begin{thm}
For any $\ell\in \mathbb{N}$, constant $\vec{\omega}\in \ES{R}{3}{1}$ and any $\tau\in \mathbb{R}$, the transition matrix $\mat{G}(\ell,\tau)$
is an orthogonal, invertible  and symplectic transformation with $2\ell$-th order accuracy, i.e.,
\begin{equation}\label{eq-Gq-1st-sp}
\inv{\mat{G}(\ell,\tau)} = \mat{G}(\ell,-\tau) = \trsp{\mat{G}(\ell,\tau)}
\end{equation}
for the LTI-QKDE.
\end{thm}

\noindent \textsc{Proof}: For a time-invariant  vector $\vec{\omega}$, the function $ \delta(\ell,\vec{\omega},\tau)$ is an odd function of $\tau$. With the help of $ \hat{\mat{\Omega}}^2 = -\mat{I}$ and $\trsp{\hat{\mat{\Omega}}} = -\hat{\mat{\Omega}}$, we immediately obtain
\begin{align*}
\trsp{\mat{G}(\ell,\tau)}
&= \trsp{[\cos \delta(\ell,\vec{\omega},\tau)\mat{I} +\sin \delta(\ell,\vec{\omega},\tau)\hat{\mat{\Omega}}]}\\
&= \cos (-\delta(\ell,\vec{\omega},\tau))\mat{I} +\sin (-\delta(\ell,\vec{\omega},\tau))\hat{\mat{\Omega}}\\
&= \mat{G}(\ell,-\tau),
\end{align*}
which implies that the transition matrix is invertible. Moreover, $\mat{G}(\ell,\tau)$ is orthogonal by
Lemma \ref{lem-Cayley-Euler}. In consequence, $\trsp{\mat{G}(\ell,\tau)} = \mat{G}(\ell,-\tau) = \inv{\mat{G}(\ell,\tau)}$.

Since the $\mat{G}(\ell,\tau)$ is constructed from the Pad\'{e} approximation $\scrd{g}{DPA}^\ell(\cdot)$ directly according to Theorem \ref{SDS-2L}, it must be a SDS with $2\ell$-th  order accuracy. Q.E.D.

\subsection{$2\ell$-th order SDS for LTV-QKDE}

For the time-varying vector $\vec{\omega}(t)$,  the LTV-QKDE is a non-autonomous Hamiltonian system essentially and the corresponding SGA  can be obtained via the concept of extended phase space and time-centered difference scheme  (See \cite{Feng2010}, Chapter 5). 
Let $\bar{t}_k = t[k] +\tau/2,
\vec{\omega}_k =\vec{\omega}(\bar{t}_k),
\mat{\Omega}_k = \mat{\Omega}(\vec{\omega}_k)$ with the similar procedure as we do
in finding the transition matrix for SDS of the LTI-QKDE.
However, we have to treat the discrete time sequence $t_0, t_0 + \frac{\tau}{2}, t_0 + \frac{3\tau}{2}, \cdots, t_0 + \frac{(2k+1)\tau}{2}, t_0 + \frac{(2k+3)\tau}{2}, \cdots$ and the corresponding symplectic transition matrices $\mat{G}^{k+1}_k(\cdot)$ at these discrete time. Thus we have to treat the initial condition $\quat{q}(t_0)$ carefully since the length of interval $[t_0, t_0 + \frac{\tau}{2}]$ is not equal to that of $[t_0 + \frac{(2k+1)\tau}{2}, t_0 + \frac{(2k+3)\tau}{2}]$. In order to avoid this phenomena and the fractional sampling problem, we use the following modified version of the discrete  vectors $\vec{\omega}_k$ and matrices $\mat{\Omega}_k$
\begin{equation} \label{eq-omegaA-LTV}
\vec{\omega}_k =\vec{\omega}(t[k]), \quad \mat{\Omega}_k = \mat{\Omega}(\vec{\omega}_k),
\end{equation}
thus we can obtain the $k$-th transition matrix at time $t[k]$  
\begin{equation}
\begin{split}
\mat{G}^{k+1}_k(\ell, \tau) 
&= g_{\ell}\left(\tau\mat{\Omega}_k/2\right)
= \phi\left(\beta(\ell,c_k)\tau\mat{\Omega}_k/2 \right)
\end{split}
\end{equation}
for the LTV-QKDE. By Lemma \ref{lem-Cayley-Euler}, it is easy to prove the following theorem.
\begin{thm} \label{Th-G-LTV}
For the LTV-QKDE~$\dfrac{\dif \quat{q}}{\dif t} = \frac{1}{2}\mat{\Omega}[\vec{\omega}(t)]\quat{q}$, let 
\begin{equation} \label{eq-c-beta-alpha}
\left\{
\begin{aligned}
c_k &= \tau^2\norm{\vec{\omega}_k}^2/4 \\
\beta_k &= \beta(\ell,c_k) \\
\alpha^\ell_k &= \alpha(\ell,c_k) = c_k \beta_k^2
\end{aligned}
\right.
\end{equation}
then the transition matrices $\mat{G}^{k+1}_k(\ell,\tau): \quat{q}[k]\mapsto \quat{q}[k+1]$ can be expressed by
\begin{equation} \label{eq-G-l-tau-c-beta-alpha}
\begin{split}
&\mat{G}^{k+1}_k(\ell, \tau)\\
&= \frac{1}{1+\alpha(\ell,c_k)}\left[(1-\alpha(\ell,c_k)\mat{I} + \tau \beta(\ell,c_k)\mat{\Omega}_k\right]\\
&= \cos [\delta\left(\ell, \vec{\omega}_k,\tau\right) ] \mat{I}
  +\sin [\delta\left(\ell, \vec{\omega}_k,\tau\right)] \hat{\mat{\Omega}}(\vec{\omega}_k),
\end{split}
\end{equation}
which generates a  SDS with $2\ell$-th order accuracy for $k= 0, 1, 2, \cdots$.
\end{thm}

Note that  $\mat{G}^{k+1}_k(\ell, \tau)$ is also  symplectic and orthogonal. However, 
$\mat{G}^{k+1}_k(\ell, -\tau) \neq \inv{[\mat{G}^{k+1}_k(\ell, \tau)]}$ since $\vec{\omega}(t)$ is time-dependent and $\vec{\omega}(t[k]-\tau) \neq \vec{\omega}(t[k]+\tau)$ in general.

\section{Symplectic Geometric Algorithms for QKDE} \label{sec-EoESGA}

\subsection{Algorithms for Computing $\beta(\ell,c)$}

For the practical computation of $\beta(\ell,c)$, it is necessary to find $\eta^\ell_k$ frequently. For the purpose of clearity, we design a simple procedure $\textsc{Eta}$ to compute $\eta^\ell_k$ according to \eqref{eq-eta-l-r}, see \Algr \ref{alg-Eta}. 
 
\begin{breakablealgorithm} \label{alg-Eta}
\caption{Compute the Parameter $\eta^\ell_k$}
\begin{algorithmic}[1]
  \Require Parameter $\ell\in \mathbb{N}$ for order and integer $k$.
  \Ensure  The value of $\eta^\ell_k$. 
\Function{Eta}{$\ell,k$}
  \State  $\eta^\ell_k \gets \dfrac{\ell-k}{(2\ell - k)(k+1)}$;
  \State \Return $\eta^\ell_k$;
\EndFunction
\end{algorithmic}
\end{breakablealgorithm}

With the help of \Algr \ref{alg-Polynomial} in Appendix \ref{appendix-poly}, we can design two fast and stable iterative algorithms for generating $\beta(\ell,c)$  according to \eqref{eq-beta-l-c-def}, see \Algr \ref{alg-P-FSIA} and \Algr \ref{alg-A-FSIA}.

\begin{breakablealgorithm}
\caption{Parallel Fast and Stable Iterative Algorithm for Generating $\beta(\ell,c)$}
\label{alg-P-FSIA}
\begin{algorithmic}[1]
\Require Positive integer $\ell$ and auxiliary parameter $c$. 
\Ensure Key auxiliary parameter $\beta$ 
\Function{PFsiaGenBeta}{$\ell, c$}
\State  $s_1 \gets \mfloor{(\ell-1)/2}, s_2 \gets \mfloor{\ell/2}$;
\State Allocate memory for  $s_1+1$ coefficients $a^\ell_j$.
\State  Allocate memory for  $s_2+1$ coefficients $b^\ell_j$.
\State $a^\ell_0 \gets \frac{1}{2}$;
\For{($j \gets 0 ; j \le s_1-1; j \gets j+1$)}
\State $a^\ell_{j+1} \gets a^\ell_j \cdot\textsc{Eta}(\ell, 2j+1)\cdot\textsc{Eta}(\ell, 2j+2)$;
\EndFor
\State $b^\ell_0 \gets 1$;
\For{($j \gets 0; j \le s_2-1; j \gets j+1$)}
\State $b^\ell_{j+1} \gets b^\ell_j \cdot \textsc{Eta}(\ell,2j)\cdot \textsc{Eta}(\ell, 2j+1)$;
\EndFor 
\State $n \gets \textsc{Polynomial}(a^\ell_0, \cdots, a^\ell_{s_1},-c)$;
\State $d \gets \textsc{Polynomial}(b^\ell_0, \cdots, b^\ell_{s_2},-c)$;
\State $\beta \gets n/d$;
\State \Return $\beta$;
\EndFunction
\end{algorithmic}
\end{breakablealgorithm}

\begin{breakablealgorithm}
\caption{Alternative Fast and Stable Iterative Algorithm for Generating $\beta(\ell,c)$}
\label{alg-A-FSIA}
\begin{algorithmic}[1]
\Require Positive integer $\ell$ and auxiliary parameter $c$. 
\Ensure Key auxiliary parameter $\beta$ 
\Function{AFsiaGenBeta}{$\ell, c$}
\State $s_1 \gets \mfloor{(\ell-1)/2}, s_2 \gets \mfloor{\ell/2}$;
\State Allocate memory for  $s_1+1$ coefficients $a^\ell_j$.
\State  Allocate memory for  $s_2+1$ coefficients $b^\ell_j$.
\State Set $b^\ell_0 \gets 1, a^\ell_0 \gets \frac{1}{2}$;
\For{($j \gets 0; j \le s_1-1; j \gets j+1$)}
\State $b^\ell_{j+1} \gets a^\ell_j \cdot\textsc{Eta}(\ell, 2j+1)$;
\State $a^\ell_{j+1} \gets b^\ell_{j+1} \cdot\textsc{Eta}(\ell, 2j+2)$;
\EndFor
\If{$ 2\mid \ell$} 
 $b^\ell_{s_2} \gets a^\ell_{s_1}\cdot \textsc{Eta}(\ell,2s_1+1)$;
\EndIf
\State $n \gets \textsc{Polynomial}(a^\ell_0, \cdots, a^\ell_{s_1},-c)$;
\State $d \gets \textsc{Polynomial}(b^\ell_0, \cdots, b^\ell_{s_2},-c)$;
\State $\beta \gets n/d$;
\State \Return $\beta$;
\EndFunction
\end{algorithmic}
\end{breakablealgorithm}

Note that the computation of $a^\ell_j$ and $b^\ell_j$ in \Algr \eqref{alg-P-FSIA} is based on  \eqref{eq-i-a} and
\eqref{eq-i-b}. However, \eqref{eq-ai-ab} is taken for \Algr \eqref{alg-A-FSIA} by comparison. 

\subsection{Algorithm for Computing Symplectic Transition Matrix} 

The symplectic transition matrix $\mat{G}^{k+1}_k(\ell,\tau)$ at any time $t_k$ can be computed based on
\eqref{eq-c-beta-alpha} and \eqref{eq-G-l-tau-c-beta-alpha} at time $t_k$.
 
 \begin{breakablealgorithm}\label{SpTranMat4Qkde}
\caption{Symplectic Transition Matrix for LTV-QKDE}
\begin{algorithmic}[1]
  \Require Angular velocity $\vec{\omega}\in \ES{R}{3}{1}$ at time $t_k$, order parameter $\ell$ and time step $\tau$.
  \Ensure  Symplectic transition matrix $\mat{G}$ for the LTV-QKDE $\fracode{\quat{q}}{t} = \frac{1}{2}\mat{\Omega}(\vec{\omega}(t))\quat{q}$ at time $t$ with $2\ell$-th order accuracy. 
\Function{SpTranMatQkdeLTV}{$\ell, \tau, \vec{\omega}$}
  \State Set matrix $\mat{\Omega}$ with $\vec{\omega}=\vec{\omega}(t_k)$:
      $$
      \mat{\Omega}
      \gets\begin{bmatrix}
       0 & -\omega_1 & -\omega_2  & -\omega_3 \\
       \omega_1 & 0 & \omega_3 & -\omega_2 \\
       \omega_2 & -\omega_3 & 0 & \omega_1 \\
       \omega_3 & \omega_2 & -\omega_1 & 0      
      \end{bmatrix}
      $$
  \State $c \gets \tau^2\norm{\vec{\omega}}^2/4$; // by \eqref{eq-c-beta-alpha}
  \State $\beta \gets \textsc{AFsiaGenBeta}(\ell,c)$; // by \eqref{eq-c-beta-alpha}
  \State $\alpha \gets c \beta^2$; // by \eqref{eq-c-beta-alpha}
  \State $\mat{G}(\ell, \tau) \gets \dfrac{1}{1+\alpha}\left[ (1-\alpha)\mat{I} + \tau\beta \mat{\Omega}\right]$; // by \eqref{eq-G-l-tau-c-beta-alpha}
  \State \Return $\mat{G}(\ell, \tau)$;
\EndFunction
\end{algorithmic}
\end{breakablealgorithm}

It is easy to find that in the Step 4 for computing the $\beta(\ell,c)$, the parallel algorithm 
\textsc{PFsiaGebBeta} described in \Algr \ref{alg-P-FSIA} is also feasible.

\subsection{Algorithm for Solving LTI-QKDE}
The $2\ell$-th order ESGA for LTI-QKDE based on Pad\'{e} approximation, $\textsc{EoEsgaQkdeLTI}$ for 
brevity\footnote{``Eo" means \textit{even order} since $2\ell$ must be an even integer. We use \textsc{EoEsgaQkdeLTI} instead of \textsc{EoEsgaLtiQkde} since the \textsc{EoEsgaQkdeLTI} and its counterpart  \textsc{EoEsgaQkdeLTV} can be merged into \textsc{EoEsgaQkde} with function overloading in object-oriented programming.},  is given in \Algr \ref{Sp-A-QKDE} by the explicit expression of $\mat{G}(\ell, \tau)$  via \eqref{eq-mat-G-ell-tau}. 
Particularly, if $\ell = 1$, then the $\mat{G}(1, \tau)$ is the same with its counterpart in the ESGA-I algorithm proposed in \cite{Zhang2018QKDE} since  $\beta(1,c) \equiv \frac{1}{2}$ for any positive number $c$.

\begin{breakablealgorithm} \label{Sp-A-QKDE}
\caption{$2\ell$-th Order Explicit Symplectic Geometric Algorithm for LTI-QKDE}
\begin{algorithmic}[1]
  \Require The constant angular velocity vector $\vec{\omega}\in \ES{R}{3}{1}$, time step $\tau$, order parameter $\ell$, time interval $[t_0, t_f]$  
  and the initial state quaternion $\quat{q}_0 $ 
  at time $t_0$.
  \Ensure  Numerical solution to the LTI-QKDE
     $\fracode{\quat{q}}{t} = \frac{1}{2}\mat{\Omega}(\vec{\omega})\quat{q}$ for $ t_0 \le t \le t_f$ with $2\ell$-th order SDS. 
\Function{EoEsgaQkdeLTI}{$\ell, \tau, \vec{\omega}, \quat{q}_0, t_0$}
  \State $t[0] \gets t_0$,
  $\quat{q}[0] \gets \quat{q}_0$;
  \State $\mat{G}(\ell,\tau) = \textsc{SpTranMatQkdeLTV}(\ell, \tau, \vec{\omega})$
  \For{($ k\gets 0; t[k]< t_f; k \gets k+1$)}
  \State $\quat{q}[k+1] \gets \mat{G}(\ell, \tau) \quat{q}[k]$;
  \State $t[k+1] \gets t[k] + \tau$;
  \EndFor
  \State \Return The sequence $\set{\quat{q}[k]}$;
\EndFunction
\end{algorithmic}
\end{breakablealgorithm}
Note that for computing $\beta(\ell,c)$ in Step 6 of \Algr \ref{Sp-A-QKDE},  we can use $\textsc{PFsiaGenBeta}(\ell,c)$ or $\textsc{AFsiaGenBeta}$ for $\textsc{FsiaGenBeta}$ according to the practical engineering requirements. Moreover, if a fixed $\ell$ is preferred, say $\ell = 4$, we can choose a specific polynomial from Table I. This strategy also holds for \Algr \ref{Sp-NA-QKDE} in the following subsection.

We remark that for real-time applications, it is not necessary to store the sequences $\set{t[k]}$ and $\set{\quat{q}[k]}$. It is enough for us to store the current state $\quat{q}[k]$ at time $t[k]$, which implies that we just need to store two real numbers.  Obviously, this remark holds for the \Algr \ref{Sp-NA-QKDE} \textsc{EoEsgaQkdeLTV} in the following subsection.

\subsection{Algorithm for Solving LTV-QKDE}

The even order ESGA for LTV-QKDE based on Pad\'{e} approximation, \textsc{EoEsgaQkdeLTV} for brevity, is presented in \Algr \ref{Sp-NA-QKDE}. It should be noted that we can share some memories for the
time-dependent parameters $\vec{\omega}_k, \mat{\Omega}_k, c_k, \beta^\ell_k, \alpha^\ell_k$ and $\mat{G}^{k+1}_k(\ell,\tau)$ since they are local variables for omputing the sequence $\set{\quat{q}[k]}$.

By comparison, \textsc{EoEsgaQkdeLTV} differs the \textsc{EoEsgaQkdeLTI} from two points: the angular velocity at time $t[k]$ should be computed and the computation of $\mat{G}$ is moved into the loop.

\begin{breakablealgorithm} \label{Sp-NA-QKDE}
\caption{$2\ell$-th Order Explicit Symplectic Geometric Algorithm for LTV-QKDE}
\begin{algorithmic}[1]
  \Require The time-varying vector $\vec{\omega}(t)\in \ES{R}{3}{1}$, positive integer $\ell$, time interval $[t_0,t_f]$, initial quaternion
 $\quat{q}_0$  and time step $\tau$.
  \Ensure  Numeric solution $\set{\quat{q}[k]}$ to the LTV-QKDE
     $\fracode{\quat{q}}{t} = \frac{1}{2}\mat{\Omega}(\vec{\omega}(t))\quat{q}$ for $ t_0\le t\le t_f$
           with $2\ell$-th order accuracy. 
\Function{EoEsgaQkdeLTV}{$\ell, \tau, \vec{\omega}(t), \quat{q}_0,t_0, t_f$}
\State $t[0] \gets t_0,
  \quat{q}[0] \gets \quat{q}_0$;
\For{($k \gets 0; t[k]< t_f; k\gets k+1 $)}
  \State $\vec{\omega}_k \gets \trsp{[\omega_1(t[k]), \omega_2(t[k]), \omega_3(t[k])]}$;
  \State $\mat{G}^{k+1}_k(\ell,\tau) \gets \textsc{SpTranMatQkdeLTV}(\ell, \tau, \vec{\omega}_k)$;
  \State $\quat{q}[k+1] \gets \mat{G}^{k+1}_k(\ell, \tau) \quat{q}[k]$;
  \State $t[k+1] \gets t[k] + \tau$;
 \EndFor
 \State \Return The sequence $\set{\quat{q}[k]}$;
\EndFunction
\end{algorithmic}
\end{breakablealgorithm}

Note that for the steps in the loop, we can use the notations 
$\vec{\omega}_k$ and $\mat{G}^{k+1}_k$ for understanding (see Theorem \ref{Th-G-LTV}). For the implementation of algorithm $\textsc{EoEsgaQkdeLTV}$, we can use
the notations $\vec{\omega}$ and $\mat{G}$ in order to share memories for the tempary (local) variables in the sense of computer programming. 

\subsection{Analysis of Computational Complexity}

Let
 $\tic{*}{(\texttt{expr})}$ and
 $\tic{+}{(\texttt{expr})} $
be the times of multiplication and addition in some operation expression $\texttt{expr}$. The \textit{time  complexity vector of computation} (TCVC)  for $\texttt{expr}$ is defined by
\begin{equation}
\timer(\texttt{expr}) = [\tic{*}{(\texttt{expr})}, \tic{+}{(\texttt{expr})}].
\end{equation}
Note that we just list two components of $\timer(\texttt{expr})$ here since for the problems and algorithms which contain massive matrix-vector operations, the multiplication and addition operations for real numbers are fundamental and essential.

Similarly, we use $T(\texttt{expr})$ to denote the computation time for $\texttt{expr}$. We also use $T_*(\texttt{expr})$ and $T_+(\texttt{expr})$ to represent the computation time for the multiplications and additions involved in $\texttt{expr}$.
Suppose that the time units for multiplication and addition are  $u_1$ and $u_2$ respectively. Let
\begin{equation}
\vec{u} = [u_1, u_2]
\end{equation}
be the vector of time units. Then the time for computing $\texttt{expr}$ will be
\begin{equation}
\begin{split}
T(\texttt{expr})
& = \braket{\timer(\texttt{expr})}{\vec{u}} \\
&= \tic{*}{(\texttt{expr})} u_1 + \tic{+}{(\texttt{expr})}u_2\\ 
&= T_*(\texttt{expr}) + T_+(\texttt{expr})
\end{split}
\end{equation}
if only multiplication and addition are essential for the total time consumed.

For an algorithm named with $\textsc{Alg}$, its TCVC is defined by
\begin{equation}
\begin{split}
\tic{}{(\textsc{Alg})} 
&= \sum_{\texttt{expr}\in \textsc{Alg}} \tic{}{(\texttt{expr})}\\
&=[\tic{*}{(\textsc{Alg})},\tic{+}{(\textsc{Alg})} ].
\end{split}
\end{equation}
The time needed for  algorithm \textsc{Alg} can be measured by
\begin{equation}
T(\textsc{Alg}) = \braket{\tic{}{(\textsc{Alg})}}{\vec{u}}= \tic{*}{(\textsc{Alg})} u_1 + \tic{+}{(\textsc{Alg})}u_2
\end{equation}
theoretically with acceptable accuracy.

It is easy to check that the TCVC for Algorithm \ref{alg-Eta} for computing $\eta^\ell_k= \textsc{Eta}(\ell,k)$ is 
\begin{equation}
\tic{}{(\textsc{Eta})} 
 = \sum_{ \texttt{expr}\in \textsc{Eta} }
 \tic{}{(\texttt{expr})} =  [3, 3].
\end{equation} 
The TCVC for computing the polynomial $\displaystyle p_s(x) = \sum^s_{i=0}c_i x^i$ via Horne's rule is
\begin{equation}
\tic{}{(\textsc{Polynomial})} 
 = \sum_{ \texttt{expr}\in \textsc{Polynomial} }
 \tic{}{(\texttt{expr})} =  [s, s]
\end{equation}
Simple algebraic operations show that The TCVC for \Algr \ref{alg-P-FSIA} and \Algr \ref{alg-A-FSIA} are 
\begin{equation}
\begin{split}
&\tic{}{(\textsc{PFsiaGenBeta})}  \\
 &= \sum_{ \texttt{expr}\in \textsc{PFsiaGenBeta} }
 \tic{}{(\texttt{expr})}  \\
 &= s_1[11,9] + s_2[11,8] + [5,3]  \\
 &= [11(s_1 + s_2)+5, 9s_1 + 8s_2 + 3]  \\
&=\left\{
     \begin{array}{ll}
     \left[11\ell-6, (17\ell-11)/2 \right],   &  \mathrm{for}~2\nmid \ell;\\
     \left[11\ell-6, (17\ell-12)/2 \right],   &  \mathrm{for}~2 \mid \ell;
     \end{array}
     \right.\\
&= [11\ell -6, (17\ell -12 + \mod(\ell,2))/2], \quad \forall \ell \in \mathbb{N}
\end{split}
\end{equation} 
and
\begin{equation}
\begin{split}
\tic{}{(\textsc{AFsiaGenBeta})}  
 &= \sum_{ \texttt{expr}\in \textsc{AFsiaGenBeta} }
 \tic{}{(\texttt{expr})}  \\
&=[6\ell-1, 5\ell-3], \quad \forall \ell\in \mathbb{N}
\end{split}
\end{equation}
respectively. Obviously, the time complexity of $\textsc{AFsiaGenBeta}$ is lower than that of $\textsc{PFsiaGenBeta}$ since there is only one loop in the former. However, if we take parallel programming, the two loops in \textsc{PFsiaGenBeta} can be executed at the same time. In this way, the difference of the time complexity  will be ignorable.

The TCVC for $\textsc{SpTranMatQkde}$ can be expressed by
\begin{equation}
\begin{split}
\tic{}{(\textsc{SpTranMatQkde})}  
 &= \sum_{ \texttt{expr}\in \textsc{SpTranMatQkde} }
 \tic{}{(\texttt{expr})}  \\
&=[6\ell+29, 5\ell+6], \quad \forall \ell\in \mathbb{N}.
\end{split}
\end{equation}
This implies that the computational cost is constant for fixed $\ell$ and varies with $\ell$ linearly when $\ell$ increases. 
  
\begin{table} 
\centering
\caption{Time and Space Consumptions of \textsc{EoEsgaQkdeLTV}}
\begin{tabular}{ccl}
\hline
Step & TCVC  & Numbers/matrices to be stored\\
\hline\hline
0    &  ---  & $t_0, t_f, \tau \in \mathbb{R}, \quat{q}[0]\in \ES{R}{4}{1}, \ell\in \mathbb{N}$ \\
1    &  ---       &  --- \\
2    &  ---      &  $k\in \mathbb{Z}, t[k]\in \mathbb{R}$ \\
3    &  ---      & $\vec{\omega}\in \ES{R}{3}{1}$ \\
4    & $ [6\ell+29, 5\ell+6]$  &  $c, \beta, \alpha\in \mathbb{R}, \mat{G}\in \ES{R}{4}{4}$\\
5    & $ [16,12] $  &  $\quat{q}[k+1]\in \ES{R}{4}{1}$\\
6   & $[0,1]$ & ---\\
7   & --- & ---\\
8   & --- & ---\\
\hline
\end{tabular}\label{Tab-TSC-QkdeLTV}
\end{table}

Let 
\begin{equation}
n = \mfloor{(t_f-t_0)/\tau},
\end{equation}
then there are $n$ iterations for the loops in \Algr \ref{Sp-A-QKDE} and  \Algr \ref{Sp-NA-QKDE}. \Tab \ref{Tab-TSC-QkdeLTV} shows that the 
TCVC for \textsc{EoEsgaQkdeLTV} is
\begin{equation}
\begin{split}
&\tic{}{(\textsc{EoEsgaQkdeLTV})}\\
&=\sum_{\texttt{expr}\in \textsc{EoEsgaQkdeLTV}} \tic{}{(\texttt{expr})}\\
&= [6\ell n + 45n, 5\ell n + 19n]
\end{split}
\end{equation} 
and the total time consumed (without optimization) can be measured by 
\begin{equation}
\begin{split}
&T(\textsc{EoEsgaQkdeLTV})\\
&=\braket{\tic{}{(\textsc{EoEsgaQkdeLTV})}}{\vec{u}}\\
&= (6\ell n +45n )u_1 + (5\ell n+19n)u_2\\
&= \mathcal{O}(n)
\end{split}
\end{equation} 
for fixed order parameter $\ell$, time units $u_1$ and $u_2$. 
By comparison, the TCVC for \textsc{EoEsgaQkdeLTI} is
\begin{equation}
\begin{split}
&\tic{}{(\textsc{EoEsgaQkdeLTI})}\\
&=\sum_{\texttt{expr}\in \textsc{EoEsgaQkdeLTI}} \tic{}{(\texttt{expr})}\\
&= [16n+6\ell+29, 13n+ 5\ell+6]
\end{split}
\end{equation} 
and the time consumed will be
\begin{equation}
\begin{split}
&T(\textsc{EoEsgaQkdeLTI})\\
&=\braket{\tic{}{(\textsc{EoEsgaQkdeLTI})}}{\vec{u}}\\
&= (16n+6\ell+29)u_1 + (13n+ 5\ell+6)u_2\\
&= \mathcal{O}(n).
\end{split}
\end{equation} 
Therefore, the time complexity for both $\textsc{EoEsgaQkdeLTI}$ and $\textsc{EoEsgaQkdeLTV}$ are linear.
Moreover, it is necessary to store the global variables $t_0, t_f, \tau$ and local variables $\omega_1, \omega_2, \omega_3,  \mat{\Omega},  \alpha, \beta, \mat{G}, k$ and $\quat{q}[k]$. The storage sharing mechanics implies that  only $45$ real numbers as well as a pointer to $\vec{\omega}(t)$ should be stored in this algorithm for real-time applications. Obviously, the space complexity is  constant, i.e., $\mathcal{O}(1)$.

We remark that we can set $\beta(\ell,c)$ according to the \Tab \ref{Tab-beta-Taylor} without invoking the \Algr \ref{alg-P-FSIA} or \Algr \ref{alg-A-FSIA}. By this way,  the time  complexity of \Algr \ref{Sp-A-QKDE} and \Algr \ref{Sp-NA-QKDE} will still be linear.

\section{Performance Evaluation}\label{sec-ValiVeri}

The fundamental aspects of verification and validation of ESGA for QKDE are discussed in detail in \cite{Zhang2018QKDE}. Our emphasis here lies in the following issues:
\begin{itemize}
\item Analysis of accuracy of $\textsc{EoEsgaQkdeLTI}$;
\item Verification and evaluatation of  $\textsc{EoEsgaQkdeLTV}$;
\begin{itemize}
\item Comparison of the NS and AS for some special time-varying $\vec{\omega}(t)$ and $\quat{q}[0]$;
\item Verification of $\textsc{EoEsgaQkdeLTV}$ with MATLAB Simulink for general case.
\end{itemize}
\end{itemize}

\subsection{Measure of NS and AS}

Let $\mat{G}_{\mathrm{AS}}(k,\tau)$ and $\mat{G}_{\mathrm{NS}}(k,\tau)$ be the AS and NS to the single step transition matrix at time $t[k]$ respectively. 
Obviously, we have
\begin{equation}
\mat{G}_{\mathrm{AS}}(k,\tau) = \mat{G}^{k+1}_k(\tau), \quad\mat{G}_{\mathrm{NS}}(k,\tau) = \mat{G}^{k+1}_k(\ell,\tau).
\end{equation} 
The error operator of step $k$ is defined by
\begin{equation}
\begin{split}
\mat{E}_\mathcal{G}[k]
&= \mat{G}_{\mathrm{AS}}(k,\tau) - \mat{G}_{\mathrm{NS}}(k,\tau) \\
&= \mat{G}^{k+1}_k(\tau) - \mat{G}^{k+1}_k(\ell,\tau).
\end{split}
\end{equation}
For the fixed $\tau
$ and  $\ell$, the absolute error of the NS  for the $k$-th solution   $\quat{q}[k]$ is 
\begin{equation}
\begin{split}
 E_k(\ell,\tau)
 &=  \norm{\scru{\quat{q}}{NS}[k] - \scru{\quat{q}}{AS}[k]}\\
 &=\sqrt{\sum^{3}_{i=0}|e_i^{AS}[k]-e_i^{NS}[k]|^2}.
 \end{split}
\end{equation} 
The maximum absolute error for solving $\quat{q}(t)$ on interval $[t_0, t_f]$ can be expressed by
\begin{equation}
\begin{split}
E_{max}(\ell,\tau) 
&= \max_{t[k]\in [t_0, t_f]} E_k(\ell,\tau) =  \\
&=  \max_{0\le k \le \mfloor{(t_f-t_0)/\tau}} \norm{\scru{\quat{q}}{NS}[k] - \scru{\quat{q}}{AS}[k]}.
\end{split}
\end{equation}

\subsection{Analysis of Accuracy of $\textsc{EoEsgaQkdeLTI}$ for LTI-QKDE}

For the LTI-QKDE, we can obtain the AS to 
the equation (\ref{eq-QKDE}) for constant $\vec{\omega}$. Since $\mat{\Omega}^2 = -\norm{\vec{\omega}}^2 \mat{I}$ and $\hat{\mat{\Omega}} = \mat{\Omega}/\norm{\vec{\omega}}$, Lemma \ref{lem-Euler} shows that
\begin{align*}
\quat{q}(t[k+1]) 
&= \exp\left((t[k+1]-t[k])\mat{\Omega}/2\right)\quat{q}(t[k]) \\
&= \left(\mat{I} \cos (\norm{\vec{\omega}}\tau/2)
 + \hat{\mat{\Omega}} \sin (\norm{\vec{\omega}}\tau/2)\right)\quat{q}(t[k]).
\end{align*}
Therefore, 
\begin{equation}
\quat{q}[k+1] = \mat{G}_{\mathrm{AS}}(k,\tau)\quat{q}[k]
\end{equation}
 where
\begin{equation}
\begin{split}
\mat{G}_{\mathrm{AS}}(k,\tau) 
&=\mat{G}^{k+1}_k(\tau) =\mat{G}(\tau)
= \exp\left(\tau\mat{\Omega}/2\right)\\
&= \mat{I} \cos (\norm{\vec{\omega}}\tau/2)
 + \hat{\mat{\Omega}} \sin (\norm{\vec{\omega}}\tau/2)
 \end{split}
\end{equation}
is the AS to the transition matrix for any $k$. On the other hand,
\begin{equation}
\mat{G}^{k+1}_k(\ell,\tau) = \mat{G}(\ell,\tau)
\end{equation}
for the LTI-QKDE. Consequently, 
\begin{equation}
\begin{split}
\mat{E}_\mathcal{G}
&= \mat{G}(\tau) - \mat{G}(\ell,\tau)\\
&= \mat{I}\left[ \cos \frac{\norm{\vec{\omega}}\tau}{2} -
\frac{1-\alpha(\ell,c)}{1+\alpha(\ell,c)}\right]\\
&\quad 
 + \hat{\mat{\Omega}} \left[ 
 \sin  \frac{\norm{\vec{\omega}}\tau}{2} - \frac{\norm{\vec{\omega}}\tau\beta(\ell,c)}{1+\alpha(\ell,c)}
 \right]
\end{split}
\end{equation}
for any $k$.
Let 
\begin{equation}
x = \norm{\vec{\omega}}\tau 
\end{equation}
then  $x$ must be a pure scalar parameter without any unit and we immediately have
\begin{equation}
\left\{
\begin{split}
&c = x^2/4,\\
&\beta(\ell,c)= \beta(\ell,x^2/4),\\
&\alpha(\ell,c)= c\beta(\ell,x)^2 = x^2[\beta(\ell,x^2/4)]^2/4.
\end{split}
\right.
\end{equation}
Let 
\begin{equation}
\left\{
\begin{split}
f_1(\ell,x) &= \cos \frac{x}{2} - 
\frac{1-\frac{x^2}{4}\left[\beta\left(\ell,\frac{x^2}{4}\right)\right]^2}{1+\frac{x^2}{4}\left[\beta\left(\ell,\frac{x^2}{4}\right)\right]^2},\\
f_2(\ell,x) &= \sin \frac{x}{2} - 
\frac{x\beta\left(\ell, \frac{x^2}{4}\right)}{1+\frac{x^2}{4}\left[\beta\left(\ell,\frac{x^2}{4}\right)\right]^2},
\end{split}
\right.
\end{equation}
then
\begin{equation}
\mat{E}_\mathcal{G} = f_1(\ell,x)\mat{I}
+ f_2(\ell,x)\hat{\mat{\Omega}}.
\end{equation}
Moreover,
\begin{equation}\label{eq-ErrMatNormEst}
\begin{split}
\norm{\mat{E}_\mathcal{G}} 
&=
\norm{f_1(\ell,x)\mat{I}
+ f_2(\ell,x)\hat{\mat{\Omega}}} \\
&\le \norm{f_1(\ell,x)\mat{I}}
+ \norm{f_2(\ell,x)\hat{\mat{\Omega}}}\\
&=\abs{f_1(\ell,x)}\norm{\mat{I}} +
\abs{f_2(\ell,x)} \norm{\mat{\Omega}}/\norm{\vec{\omega}}\\
&= 2(\abs{f_1(\ell,x)} + \abs{f_2(\ell,x)})\\
&=\mathcal{O}(x^{2\ell+1})\\
&=\mathcal{O}(\tau^{2\ell+1})
\end{split}
\end{equation}
for sufficiently small $x = \norm{\vec{\omega}}\tau$ and any $k$. 
According to \Tab \ref{Tab-beta-Taylor} and (\ref{eq-ErrMatNormEst}), we can obtain \Tab \ref{tab-err-G-A-QKDE}. It is clear that the upper bound $2(\abs{f_1(\ell,x)} + \abs{f_2(\ell,x)})$ for 
$\norm{\mat{E}_\mathcal{G}} $ decreases rapidly when $\ell$ increases.

Furthermore, for the LTI-QKDE we have the single step computational error 
\begin{align*}
&\scru{\quat{q}}{AS}[k+1] - \scru{\quat{q}}{NS}[k+1]\\
&=\mat{G}_{\mathrm{AS}}(\tau)\scru{\quat{q}}{AS}[k] - \mat{G}_{\mathrm{NS}}(\tau)\scru{\quat{q}}{NS}[k]\\
&=\left(\mat{G}_{\mathrm{AS}}(\tau) - \mat{G}_{\mathrm{NS}}(\tau)\right)\scru{\quat{q}}{AS}[k]+\mat{G}_{\mathrm{NS}}(\tau)\left(\scru{\quat{q}}{AS}[k]-\scru{\quat{q}}{NS}[k]\right)
\end{align*}
and its estimation
\begin{align*}
&\norm{\scru{\quat{q}}{AS}[k+1] - \scru{\quat{q}}{NS}[k+1]}\\
&\le \norm{(\mat{G}_{\mathrm{AS}}(\tau) - \mat{G}_{\mathrm{NS}}(\tau))\scru{\quat{q}}{AS}[k]} + \norm{\mat{G}_{\mathrm{NS}}(\tau)(\scru{\quat{q}}{AS}[k]-\scru{\quat{q}}{NS}[k])}\\
&\le \norm{\mat{G}_{\mathrm{AS}}(\tau) - \mat{G}_{\mathrm{NS}}(\tau)}\norm{\scru{\quat{q}}{AS}[k]}
+\norm{\mat{G}_{\mathrm{NS}}(\tau)(\scru{\quat{q}}{AS}[k]-\scru{\quat{q}}{NS}[k])}\\
&=\norm{\mat{G}_{\mathrm{AS}}(\tau) - \mat{G}_{\mathrm{NS}}(\tau)} + \norm{\scru{\quat{q}}{AS}[k]-\scru{\quat{q}}{NS}[k]}\\
&= \norm{\mat{E}_\mathcal{G}} + \norm{\scru{\quat{q}}{AS}[k]-\scru{\quat{q}}{NS}[k]}
\end{align*}
since we have $\norm{\scru{\quat{q}}{AS}[k]} \equiv 1$ for any  integer $k$ and $\mat{G}_{\mathrm{NS}}(\tau)$ is an orthogonal matrix.  For $k=0$, we have
$$
E_0(\ell,\tau) = \norm{\scru{\quat{q}}{AS}[0]-\scru{\quat{q}}{NS}[0]} = 0
$$
because of $\scru{\quat{q}}{AS}[0] =\scru{\quat{q}}{NS}[0] = \quat{q}[0]$.
In consequence, 
\begin{equation}
\begin{split}
 E_k(\ell,\tau) 
 &= \norm{\scru{\quat{q}}{AS}[k]-\scru{\quat{q}}{NS}[k]} \\
 &\le \norm{\mat{E}_\mathcal{G}} + E_{k-1}(\ell,\tau)\\
 &\le k \norm{\mat{E}_\mathcal{G}} 
 \end{split}
\end{equation}
for any $k$ by induction. Therefore, 
\begin{equation}
\begin{split}
E_{max}(\ell,\tau) 
&\le 2 k_{max} (\abs{f_1(\ell,x)} + \abs{f_2(\ell,x)})\\
& = 2 \mfloor{\frac{t_f - t_0}{\tau}} (\abs{f_1(\ell,x)} + \abs{f_2(\ell,x)}) \\
&=\mathcal{O}((t_f-t_0)\tau^{2\ell}).
\end{split}
\end{equation}
This implies that the maximum absolute error $E_{max}(\ell,\tau)$ has an upper bound with order $\mathcal{O}(\tau^{2\ell})$ for fixed value of $t_f-t_0$. 
\begin{table*}
\centering
\caption{Accuracy of $\mat{G}(\ell,\tau)$ for LTI-QKDE: $\norm{\mat{E}_\mathcal{G}} \le 2(\abs{f_1(x)} + \abs{f_2(x)})$}
\begin{spacing}{1.5}
\begin{tabular}{clll}
\hline 
 $\ell$  & $f_1(\ell,x)$  & $f_2(\ell,x)$  & $2(\abs{f_1(x)} + \abs{f_2(x)})=\mathcal{O}(x^{2\ell+1})$ \\
\hline \hline
$ 1$    & $-\dfrac{x^4}{192}+\mathcal{O}\left(x^6\right)$   &  $\dfrac{x^3}{96}+\mathcal{O}\left(x^5\right)$  &  $\dfrac{x^3}{48} + \mathcal{O}\left(x^4\right)=\mathcal{O}(x^{3})$\\
$ 2$    & $-\dfrac{x^6}{46080}+\mathcal{O}\left(x^7\right)$   &  $\dfrac{x^5}{23040}+\mathcal{O}\left(x^7\right)$  &  $\dfrac{x^5}{11520}+  \mathcal{O}\left(x^6\right)=\mathcal{O}(x^{5})$\\
$ 3$    &  $-\dfrac{x^8}{25804800}+\mathcal{O}\left(x^{10}\right)$  &  $\dfrac{x^7}{12902400}+\mathcal{O}\left(x^9\right)$  &  $\dfrac{x^7}{6451200}+\mathcal{O}\left(x^8\right)=\mathcal{O}(x^{7})$\\
$ 4$    &  $-\dfrac{x^{10}}{26011238400}+\mathcal{O}\left(x^{12}\right)$  &  $\dfrac{x^9}{13005619200}+\mathcal{O}\left(x^{11}\right)$  &  $\dfrac{x^9}{6502809600} + \mathcal{O}\left(x^{10}\right)=\mathcal{O}(x^{9})$\\
$ 5$    &  $-\dfrac{x^{12}}{41201801625600}+\mathcal{O}\left(x^{13}\right)$  &  $\dfrac{x^{11}}{20600900812800}+\mathcal{O}\left(x^{12}\right)$  &  $\dfrac{x^{11}}{10300450406400} + \mathcal{O}\left(x^{12}\right)=\mathcal{O}(x^{11})$\\
\hline
\end{tabular}
\end{spacing}
\label{tab-err-G-A-QKDE}
\end{table*}

\begin{figure*}[h]
\subfigure[Effect of $\ell$ on $E_{max}(\ell,\tau)$ for fixed time step $\tau$.]{
    \includegraphics[width=0.5\textwidth]{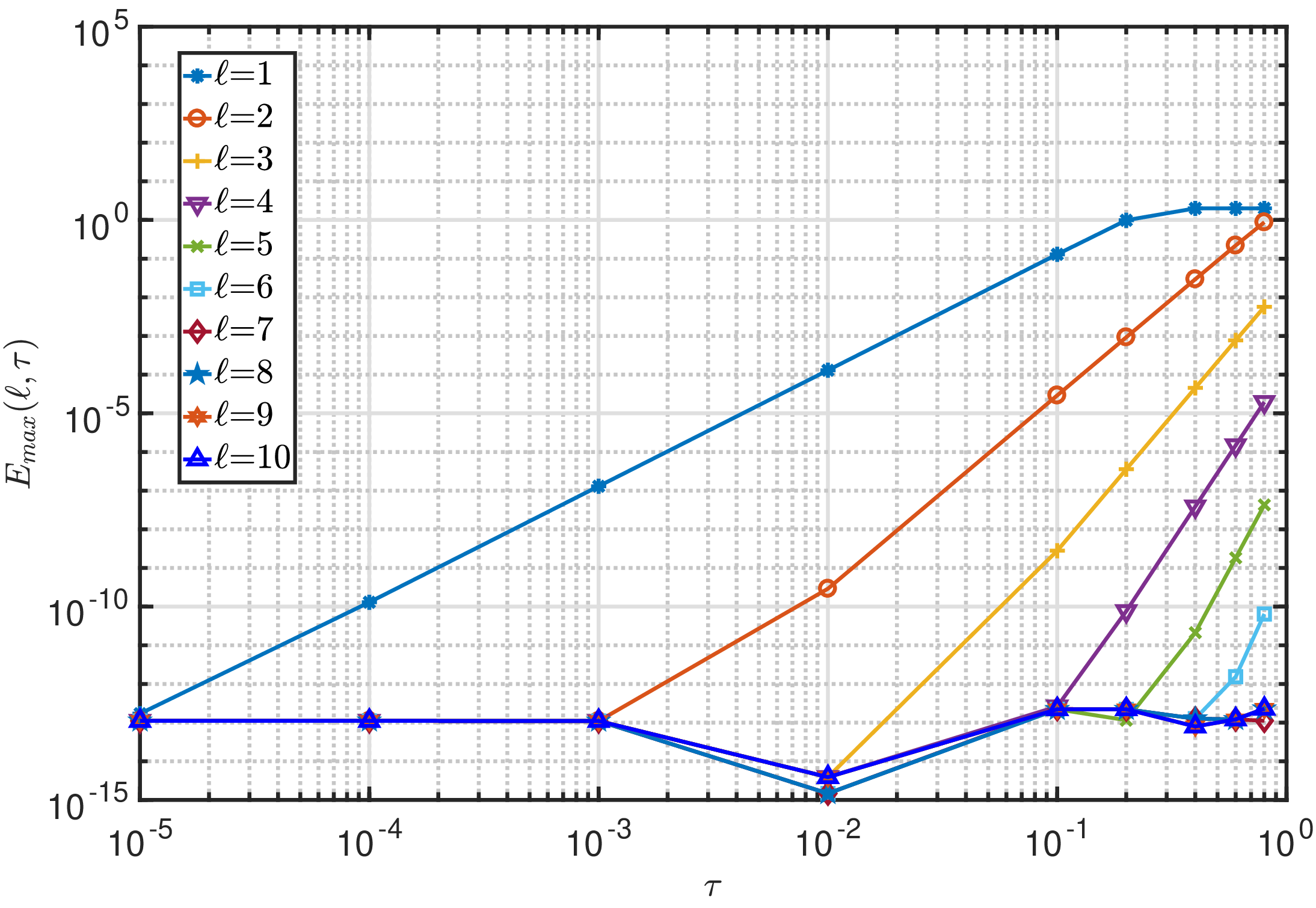}
    \label{fig-AQkde-Emax-ell}
}
\subfigure[Effect of $\tau$ on $E_{max}(\ell,\tau)$ for fixed order parameter $\ell$.]{
\includegraphics[width=0.5\textwidth]{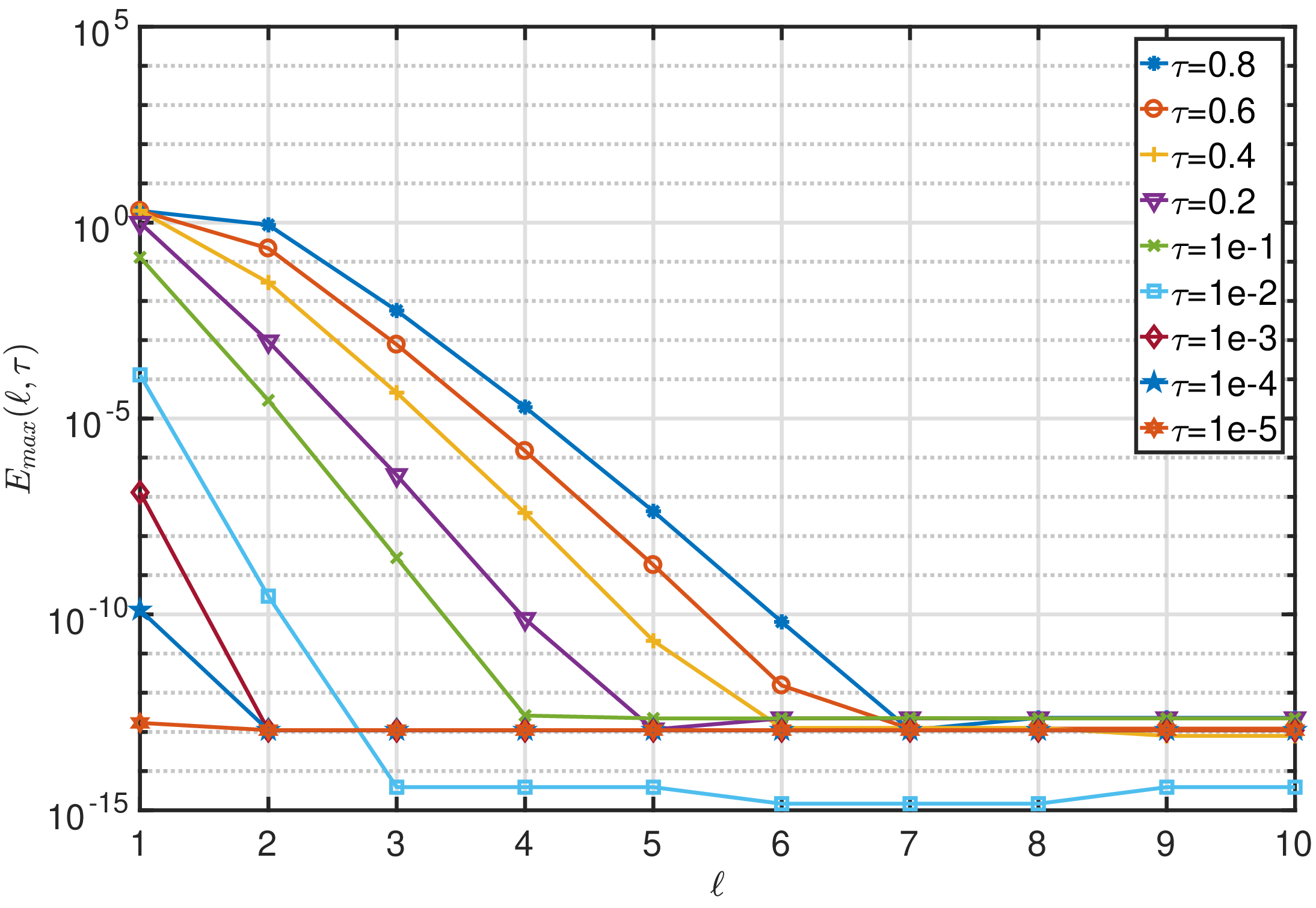}
    \label{fig-AQkde-Emax-tau}
}
\caption{$E_{max}(\ell,\tau)$ of LTI-QKDE for different order and time step with a long time span $[t_0, t_f] = [0, 2000]$, constant angular velocity $\vec{\omega} = \trsp{\left[\pi\sin \dfrac{\pi}{8}, -\dfrac{\pi}{3}\cos\dfrac{\pi}{8}, -2\sin\dfrac{\pi}{3}\right]}$   and initial state $\quat{q}_0=\trsp{[1,0,0,0]}$.}
\label{fig-AQkde-Emax}
\end{figure*}

For the order parameter $\ell \in \set{1, 2, \cdots, 10}$ and  time step 
$\tau \in \set{10^{-6}, 10^{-5}, 10^{-4}, 10^{-3}, 10^{-2},10^{-1},  0.2, 0.4, 0.6, 0.8}$ (seconds), \Fig \ref{fig-AQkde-Emax} demonstrates the impacts of $\ell$ and $\tau$ on the maximum absolute error $E_{max}(\ell,\tau)$ for the LTI-QKDE. 

\Fig \ref{fig-AQkde-Emax-ell} shows that if $\tau\le 10^{-3}$ second, we can set $\ell \ge  2$ since we have $E_{max}(\ell,\tau)\sim 10^{-13}$ for $\ell \ge 2$ instead of $E_{max}(\ell,\tau)\sim 10^{-7}$ for $\ell = 1$. If we set $E_{max}(\ell,\tau)\le 10^{-5}$, we should set $ \tau \le 0.007$ second for $\ell =1$ and $\tau \le 0.11$ for $\ell\ge 2$. Moreover, for $\ell \ge 4$, $E_{max}(\ell,\tau)$ will be small enough which implies the time step can be set large enough ($0< \tau < 1$) and the requirement for the performance of hardware will be easily satisfied. 

\subsection{Simulation of $\textsc{EoEsgaQkdeLTV}$ for LTV-QKDE}

\subsubsection{Initial Value $\quat{q}[0]$ for the QKDE}

The initial value  $\quat{q}[0] = \trsp{[e_0(t_0), e_1(t_0), e_2(t_0), e_3(t_0)]}$ for the QKDE can be determined by the initial values of the Euler angles $\psi(t_0), \theta(t_0)$ and $\phi(t_0)$, where $\psi$ is yaw angle about $x$-axis, $\theta$ is the pitch angle about $y$-axis, $\phi$ is the roll angle about $z$-axis. The Euler angles $(\psi,\theta,\phi)$ can be converted to the quaternion $\quat{q}=\trsp{[e_0,e_1,e_2,e_3]}$ according to the following formula \cite{Cooke1992,Allerton2009,Diston2009,Kim2023Rotation}
\begin{equation}
\left\{
\begin{aligned}
e_0 & = +\cos \frac{\psi}{2}\cos \frac{\theta}{2}\cos\frac{\phi}{2} + \sin \frac{\psi}{2}\sin \frac{\theta }{2}\sin\frac{\phi }{2},\\
e_1  &= +\cos \frac{\psi }{2}\cos \frac{\theta }{2}\sin\frac{\phi }{2}- \sin \frac{\psi }{2}\sin \frac{\theta }{2}\cos\frac{\phi }{2},\\
e_2  &= +\cos \frac{\psi }{2}\sin \frac{\theta }{2}\cos\frac{\phi }{2} + \sin \frac{\psi }{2}\cos \frac{\theta }{2}\sin\frac{\phi }{2},\\
e_3  &= -\cos \frac{\psi }{2}\sin \frac{\theta }{2}\sin\frac{\phi }{2} + \sin \frac{\psi }{2}\cos \frac{\theta }{2}\cos\frac{\phi }{2}.
\end{aligned}
\right.
\end{equation}
Particularly, we have the two special configurations for the Euler angle  and the quaternion, viz.
\begin{equation} \label{eq-init-state-1}
\begin{bmatrix}
\psi \\ \theta \\ \phi
\end{bmatrix} =\begin{bmatrix} 0 \\ \xi \\0
\end{bmatrix} \longleftrightarrow  \quat{q}[0] =
\begin{bmatrix}
\cos \frac{\xi}{2} \\ 0 \\ \sin \frac{\xi}{2} \\ 0\end{bmatrix} 
\end{equation}
and
\begin{equation} \label{eq-init-state-2}
\begin{bmatrix}
\psi \\ \theta \\ \phi
\end{bmatrix} =\begin{bmatrix} 0 \\ 0 \\0
\end{bmatrix} \longleftrightarrow  \quat{q}[0] =
\begin{bmatrix}
1 \\ 0 \\ 0 \\ 0\end{bmatrix} 
\end{equation}
The equation \eqref{eq-init-state-1} or \eqref{eq-init-state-2} can be used to set the initial values $\quat{q}[0]$ for solving the attitudes.

\subsubsection{A special LTV-QKDE with AS}
For the constant parameters 
 $\omega_0, \xi$, initial state $\quat{q}[0]$ specified by \eqref{eq-init-state-2} and the time-varying angular velocity
\begin{equation}  \label{eq-omega-special}
\vec{\omega}(t) = \begin{bmatrix}
-\omega_0(1-\cos \xi)\\
-\omega_0 \sin \xi \sin (\omega_0 t)\\
\omega_0 \sin \xi \cos (\omega_0 t)
\end{bmatrix},
\end{equation}
it is easy to check that the corresponding AS to the LTV-QKDE is
\begin{align*}
\scru{\quat{q}}{AS}(t) 
&= \trsp{[e_0(t),e_1(t),e_2(t),e_3(t)]}\\
&= \trsp{\left[\cos \frac{\xi}{2},0,\sin \frac{\xi}{2} \cos(\omega_0 t),
\sin \frac{\xi}{2}\sin(\omega_0 t)\right]}.
\end{align*}
For the discrete version, we have 
$$\scru{\quat{q}}{AS}[k] =\scru{\quat{q}}{AS}(t_0 + k\tau) = \mat{G}^k_{k-1}(\tau)\scru{\quat{q}}{AS}[k-1].$$ 
On the other hand, the NS to the LTV-QKDE by the \textsc{EoEsgaQkdeLTV} is
\begin{equation}
\begin{aligned}
\scru{\quat{q}}{NS}_{\ell,\tau}[k] 
&= \trsp{\left[\scru{e}{NS}_0[k],\scru{e}{NS}_1[k],\scru{e}{NS}_2[k],\scru{e}{NS}_3[k] \right]} \\
&= \mat{G}^{k}_{k-1}(\ell,\tau)\scru{\quat{q}}{NS}_{\ell,\tau}[k-1]. 
\end{aligned}
\end{equation}

\begin{figure}[h]
\resizebox{8.7cm}{6cm}{\includegraphics{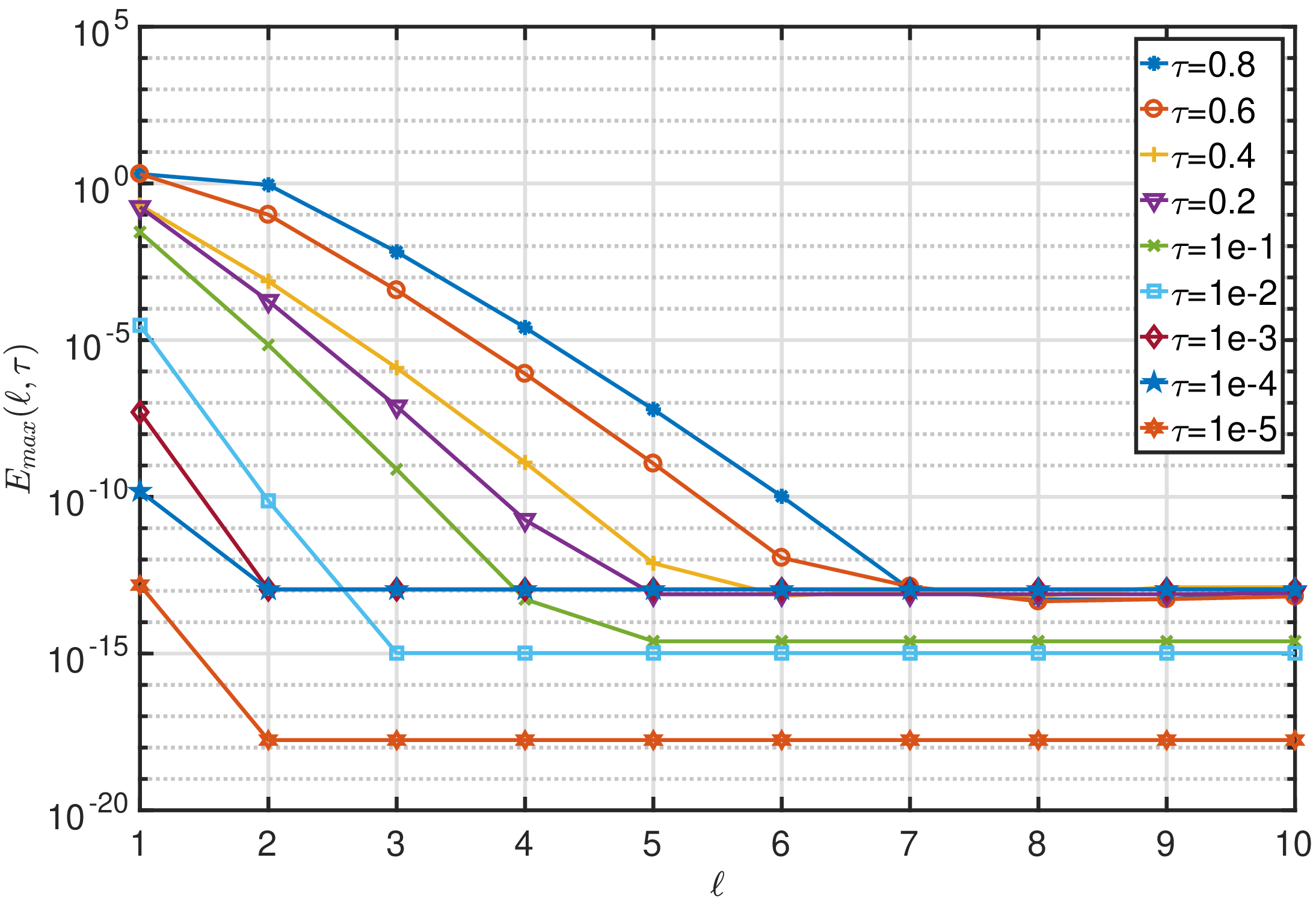}}
\caption{Effect of $\ell$ and $\tau$ on $E_{max}(\ell,\tau)$ for LTV-QKDE}
\label{fig-NaQkde-Emax}
\end{figure}

\Fig \ref{fig-NaQkde-Emax} demonstrates a numeric example of LTV-QKDE with AS. Here we set    time span $t_f - t_0 = 2000$ seconds, $\omega_0 = 2\pi$,$\xi = \pi/80$, initial state 
$\quat{q}_0=\scru{\quat{q}}{AS}(t_0)$ 
and $\vec{\omega}(t)$ according to (\ref{eq-omega-special}). Obviously, larger order parameter $\ell$ implies smaller $E_{max}(\ell,\tau)$ for fixed $\tau$.
If we set $\tau = 10^{-2}$ and need  $E_{max}(\ell,\tau)\sim 10^{-5}$, then any $\ell \ge 1$ is acceptable. 
However, if we set $\tau = 10^{-1}$ and $E_{max}(\ell,\tau)\le 10^{-5}$, then we should set $\ell \ge 2$. 
For $\ell \ge 4$ and  $E_{max}(\ell,\tau) \le 10^{-4}$,
 a big step $\tau = 0.8$ is satisfactory. The figure shows that for fixed $E_{max}(\ell,\tau)$, we can set large step for high order algorithms. In other words, with configurable order parameter $\ell$ we can choose proper hardware for practical flight control systems by setting the step $\tau$ for sampling.  Moreover, for fixed $\tau$, a larger $\ell$ means better manoeuvrability  in flight since it traces the variation of $\vec{\omega}(t)$ better. Obviously, this observation  may be valuable for designing aircraft, spacecrafts, torpedoes and so on.

\subsubsection{Simulation of LTV-QKDE with MATLAB Simulink}

It is well known that there is no AS for a LTV system such as the LTV-QKDE. In addition to the NS, simulation is another method to explore the solution to  LTV-QKDE. 

\begin{figure}[h]
\subfigure[MATLAB Simulink diagram.]{
\includegraphics[width=0.5\textwidth]{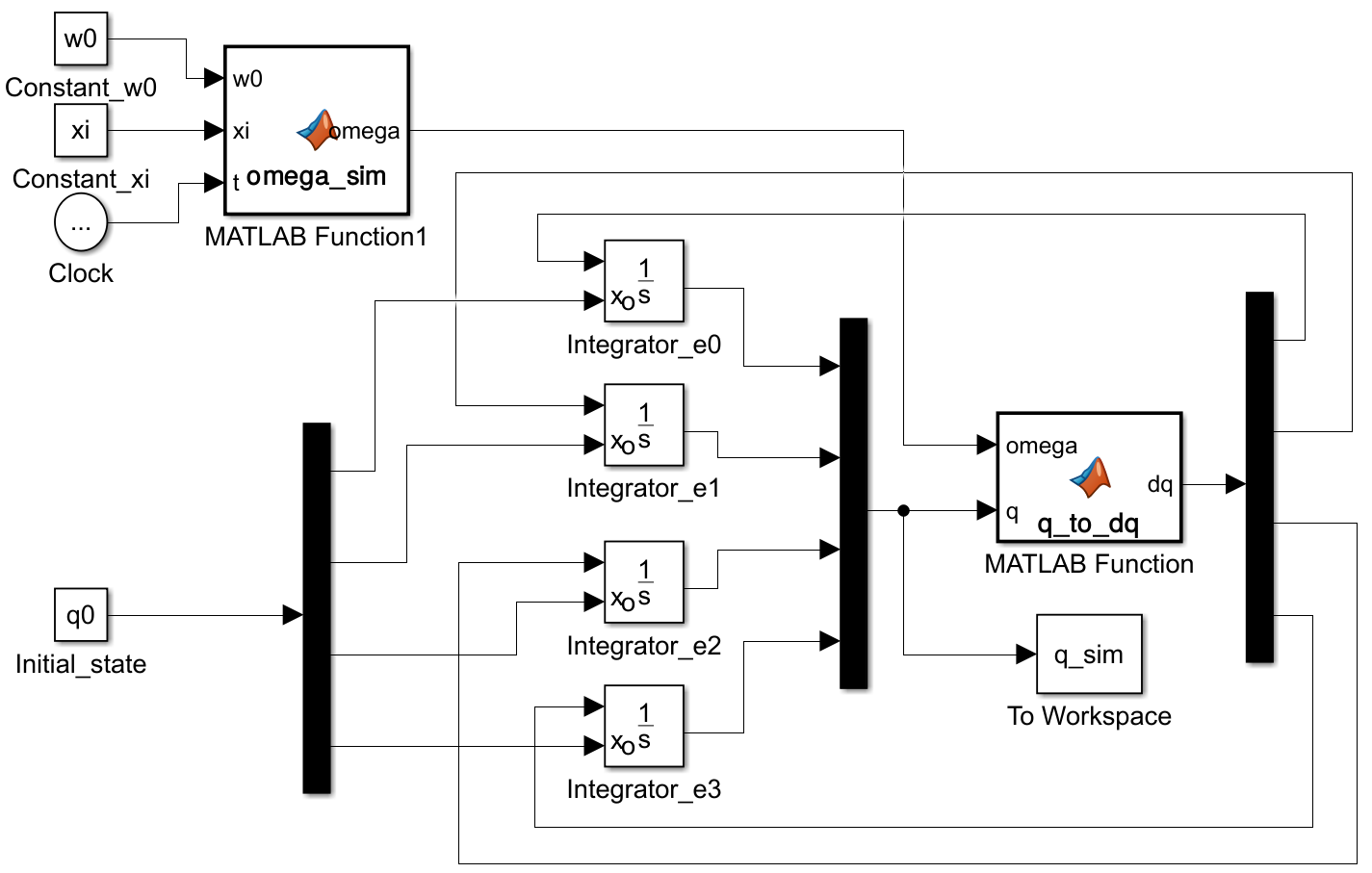}
\label{fig-NAQkde-sim-diagram}
}
\subfigure[Error between Pseudo AS to LTV-QKDE and NS with MATLAB Simulink]{
\includegraphics[width=0.5\textwidth]{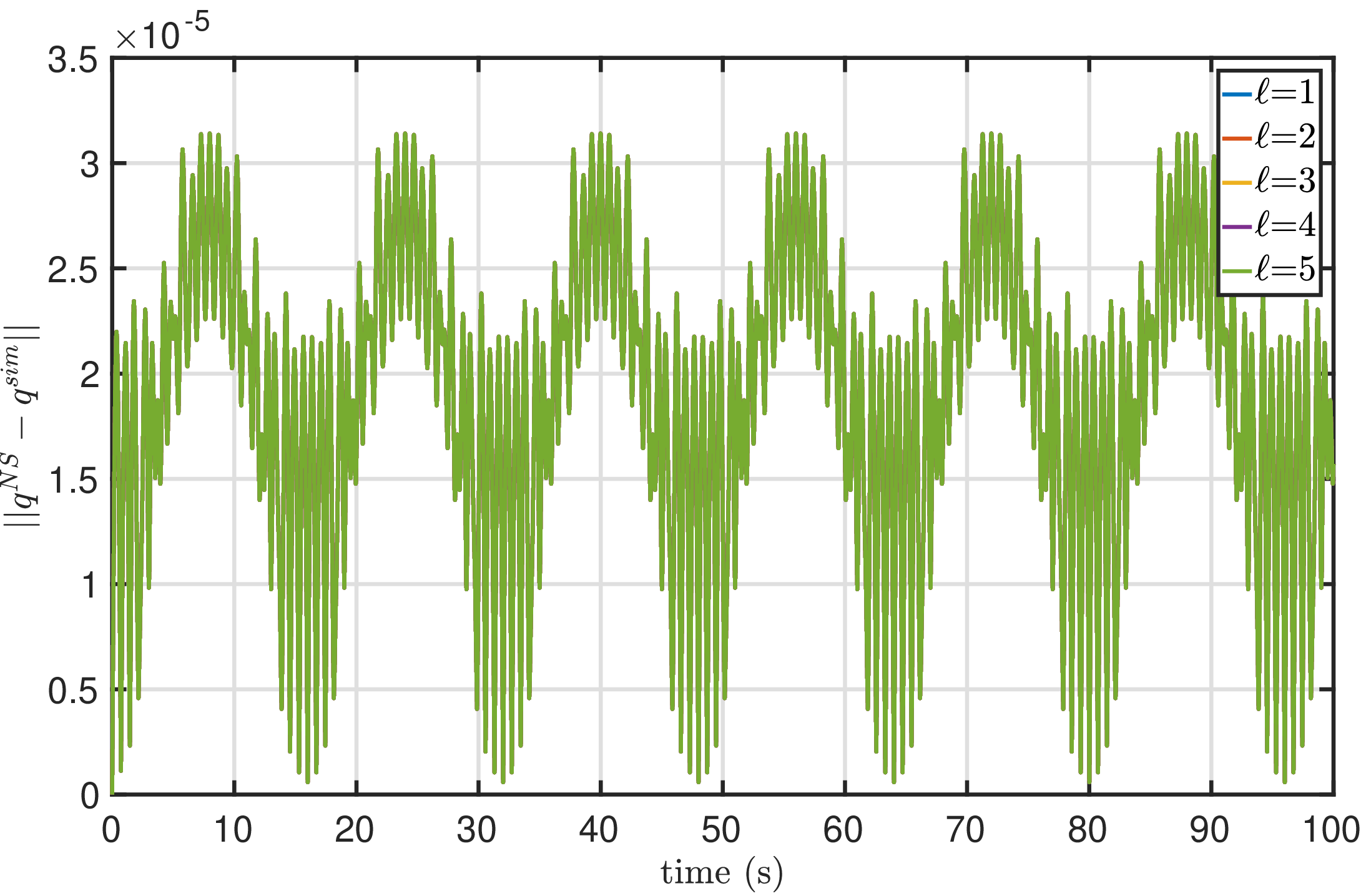}
\label{fig-NAQkde-sim-res}
}        
\caption{Simulation with MATLAB Simulink}
\label{fig-NaQkde-simulink}
\end{figure}

\Fig \ref{fig-NaQkde-simulink} illustrates the results of simulation for the LTV-QKDE.  \Fig \ref{fig-NAQkde-sim-diagram} shows the diagram for simulating LTV-QKDE. \Fig \ref{fig-NAQkde-sim-res} demonstrates the error 
between the AS obtained by Simulink and the NS obtained by our SDS. Note that the error is represented by the $\norm{\scru{\quat{q}}{sim}-\scru{\quat{q}}{NS}}$ for $\ell \in \{1,2,3,4,5\}$ with the time step $\tau = 10^{-5}$, the time span $[t_0, t_f] = [0, 100]$, the  time-varying angular velocity $\vec{\omega}(t) = \trsp{[-\omega_0\cos(\xi t)\exp(-\omega_0t),
 -\omega_0\sin(\omega_0t),
 \omega_0\cos(\xi t) \cos(\omega_0 t)]}$, and the initial state $\quat{q}_0$ determined by \eqref{eq-init-state-2} such that $\omega_0 = 2\pi$ and $\xi = \pi/80$.

\subsection{Discussion}

For the purpose of verification and validation, our emphasis is put on 
numerical simulation and system simulation since strict mathematical proofs and analysis are provided in this paper. For the readers who cares about real-world or benchmark problems, please see  \cite{Allerton2009}, \cite{Diston2009} and so on. This work is a big extension of \cite{Zhang2018QKDE}, thus the comparison of SDS and traditional difference scheme is omitted. We remark that the adaptive step size method \cite{Skvortsov2012} is not considered since it increases the complexity when implementing the SDS with hardware due to the fixed sampling frequency of IMU and it is not symplectic. 

Our work is based on the single step numerical method driven by the real-time computation and hardware implementation. The multiple steps method is not considered in this paper. The typical multiple steps method is the \textit{triple-jump composition} method popularized by  
Creutz \cite{Creutz1989}, Yosida\cite{Yoshida1990} and Suzuki \cite{Suzuki1990}. Furthermore, Yosida's symplectic method is based on the separability of the Hamiltonian which is not satisfied for the QKDE, neither Suzuki's method nor Creutz's method is symplectic. 

We also remark that the computational costs of division operation  and  multiplication are not the same in general. However, the analysis of computational cost with the time complexity vector of computation $\vec{u} =[u_{+}, u_*, u_{\div}] = [u_1, u_2, u_3]$ instead of $\vec{u} = [u_{+}, u_*] = [u_1, u_2]$ will lead to the same computational complexity. In consequence, it is enough to consider the addition and multiplication for the operations of float numbers. The essence lies in whether there are nested loops or not in the algorithms instead of the difference of division and multiplication.     

\section{Conclusions}\label{sec-Conclusion}

The essential observations for solving the QKDE with high order ESGA include four aspects: 
\begin{itemize}
\item the general LTV-QKDE can be modeled by the linear Hamiltonian system with infinitesimal symplectic structure;
\item the symplectic Pad\'{e} approximation can be used to construct the SDS for QKDE with $2\ell$-th order accuracy; 
\item the symplectic transform (single-step transition operator) can be simplified by the Pad\'{e}-Cayley lemma;
\item the parameters for the SDS can be obtained with parallel and alternative iterative methods;
\end{itemize}
The performance of algorithms --- \textsc{EoEsgaQkdeLTV} and \textsc{EoEsgaQkdeLTI} ---  are attractive due to the following reasons:
\begin{itemize}
\item there are no accumulative computational errors in the sense of long term time since they are symplectic;
\item the order parameter $\ell$ is configurable, which benefits  the designer of practical engineering systems such as aircraft, spacecrafts and so on according to the acceptable angular velocity;
\item the time  complexity of computation for the algorithms proposed   is linear while the space complexity of computation is constant, thus our algorithms are appropriate for real-time applications;
\item the maximum absolute error when solving the LTI-QKDE is  controlled by the time span $t_f-t_0$, time step $\tau$ and the order parameter $\ell$ through the relation $E_{max}(\ell,\tau) =
 \mathcal{O}((t_f-t_0)\tau^{2\ell})$, and the simulation result shows that this relation also holds for some LTV-QKDE.
\end{itemize}
Since the LTI-QKDE is a special case of LTV-QKDE and the \textsc{EoEsgaQkdeLTV} surpasses the algorithms ESGA-I and ESGA-II proposed in \cite{Zhang2018QKDE}, it is enough for us to use the \textsc{EoEsgaQkdeLTV} algorithm for solving QKDE. Generally, $\ell = 2$ is enough (which means fourth order precision) for the applications with low speed and $\ell = 5$ is enough (which means tenth order precision) for the applications with high speed.

\section*{Acknowledgement}

Our thanks go for Dr. Zi-Hao Wang for his help in simulation of LTV-QKDE with MATLAB Simulink demonstrated in \Fig \ref{fig-NaQkde-simulink}.

\section*{Code Availability Statement}

The code for the algorithms appeared in this paper
have been released on the following GitHub site: 
\begin{center}
\url{https://github.com/
GrAbsRD/EoESGA}
\end{center}

\appendix

\section{Hamiltonian System} \label{appendix-Hamilton}
W. R. Hamilton introduced the canonical differential equations \cite{Arnold1989,Arnold2007}
\begin{equation*}
\fracode{p_i}{t} = -\fracpde{H}{q_i}, \quad \fracode{q_i}{t} = \fracpde{H}{p_i}, \quad i =1, 2, \cdots, N
\end{equation*}
for problems of geometrical optics, where $p_i$ are the generalized momentum,
$q_i$ are the generalized displacements and $H = H(p_1, \cdots, p_N, q_1, \cdots, q_N)$ is the Hamiltonian, viz., the total energy of the system. Let
$\vec{p} =\trsp{[p_1, \cdots, p_N]}\in \ES{R}{N}{1}$,
$\vec{q} = \trsp{[q_1, \cdots, q_N]}\in \ES{R}{N}{1}$, and
$\vec{z} =\trsp{[p_1, \cdots, p_N, q_1, \cdots, q_N]}= \trsp{[\trsp{\vec{p}}, \trsp{\vec{q}}]}\in \ES{R}{2N}{1}$,
then $H = H(\vec{p}, \vec{q}) = H(\vec{z})$ can be specified by  $\vec{z}$ in the $2N$-dimensional phase space. Since the gradient of $H$ is
 $$
H_{\vec{z}}
=\trsp{\left[ \fracpde{H}{p_1}, \cdots, \fracpde{H}{p_N}, \fracpde{H}{q_1}, \cdots, \fracpde{H}{q_N}\right]} \in \ES{R}{2N}{1}.
$$
Then the canonical equation is equivalent to
\begin{align} \label{eq-Heq}
\fracode{\vec{z}}{t} =  \inv{\mat{J}_{2N}} \cdot H_{\vec{z}}(\vec{z}), \quad
\mat{J}_{2N} = \begin{bmatrix} \mat{O}_N & \mat{I}_N \\ -\mat{I}_N & \mat{O}_N\end{bmatrix}
\end{align}
where $\mat{I}_N$ is the $N$-by-$N$ identical matrix, $\mat{O}_N$ is the $N$-by-$N$ zero matrix and $\mat{J}_{2N}$
is the $2N$-th order standard symplectic matrix \cite{Feng1984,Feng2010}. For simplicity, the subscripts in $\mat{I}_{N}, \mat{I}_{2N}$ and $\mat{J}_{2N}$ may be omitted. Any system which can be described by  (\ref{eq-Heq}) is called an \textit{Hamiltonian system}. There are some fundamental properties for the canonical equation of Hamiltonian system \cite{Arnold1989, Kong2009, Franco2013, Hernandez2016}:
\begin{itemize}
\item[(i)] it is invariant under the symplectic transform (phase flow);
\item[(ii)] the evolution of the system is the evolution of symplectic transform;
\item[(iii)] the symplectic symmetry and the total energy of the system can be preserved simultaneously and automatically.
\end{itemize}

\section{Cayley Transform and Its Mirror} 
\label{sec-appen-cayley}

The Caylay transform is defined by \cite{Rudin1987}
\begin{equation} \label{eq-cayley-complex}
\varphi(z) = \frac{1-z}{1+z},  \quad  z\in \mathbb{C}
\end{equation}
which is introduced by Cayley original and developed by Weyl \cite{Weyl1946} in 1940s as well as by Wu \cite{WuFC2009} in 2009.
The complex variable $z\in \mathbb{C}\backslash \set{-1}$ in \eqref{eq-cayley-complex} can be generalized to the matrix form, viz. \cite{Weyl1946}
\begin{equation}
\varphi(\mat{A}) = \dfrac{\mat{I}-\mat{A}}{\mat{I}+\mat{A}}, \quad
\mat{A}\in \ES{R}{n}{n}
\end{equation}
such that 
$\mat{I}+\mat{A}\in \GrGL{n}{R}=\set{\mat{B}\in \ES{R}{n}{n}: \det(\mat{B})\neq 0}$ 
where $\dfrac{\mat{I}-\mat{A}}{\mat{I}+\mat{A}}$ denotes the multiplication $(\mat{I}-\mat{A})\inv{(\mat{I}+\mat{A})} = \inv{(\mat{I}+\mat{A})}(\mat{I}-\mat{A})$. Although there are various properties of Cayley transform \cite{Weyl1946,WuFC2009}, we just list the following theorem according to our purpose in this paper :
\begin{thm}
For the non-exceptional matrix $\mat{A}\in \ES{R}{n}{n}$, i.e. $\mat{I}+\mat{A}\in \GrGL{n}{R}$, the following statements hold:
\begin{itemize}
\item[i)] if $\lambda$ is the eigenvalue  of $\mat{A}$ such that $\mat{A}\vec{p} = \lambda\vec{p}$, then $\varphi(\lambda)$ is the eigenvalue of $\varphi(\mat{A})$, viz. $\varphi(\mat{A})\vec{p}=\varphi(\lambda)\vec{p}$;  
\item[ii)] if $\forall n\in\mathbb{N}$, then $\varphi^{2n}(\mat{A}) = \mat{A},\varphi^{2n-1}(\mat{A})=\varphi(A)$;
\item[iii)] if $\mat{A}\in \GrGL{n}{R}$, then $\varphi(\inv{\mat{A}}) = -\varphi(\mat{A})$;
\item[iv)] if $\varphi(\mat{A})\in \GrGL{n}{R}$, then $\inv{\varphi(\mat{A})}=\varphi(-\mat{A})$;
\item[v)] $\varphi(\mat{A}) = 2\inv{(\mat{I}+\mat{A})}-\mat{I} = -2\mat{A}\inv{(\mat{I}+\mat{A})}+\mat{I}$;
\end{itemize} 
\end{thm}

The matrix transform
\begin{equation}
\scrd{\varphi}{mct}(\mat{A})= \varphi(-\mat{A}),
\end{equation}
it is called the \textit{mirrored Cayley transform} of $\mat{A}$. 
Obviously, for the non-exceptional matrix 
$\mat{A}\in \ES{R}{n}{n}$
such that $\mat{I}\pm \mat{A}\in \GrGL{n}{R}$ are invertible, we have
\begin{equation}
\begin{aligned}
\scrd{\varphi}{mct}(\mat{A})
&= \dfrac{\mat{I}+\mat{A}}{\mat{I}-\mat{A}} \\
&=
(\mat{I}+\mat{A})\inv{(\mat{I}-\mat{A})} \\
&= \inv{(\mat{I}-\mat{A})}(\mat{I}+\mat{A})
\end{aligned}
\end{equation}

\section{Horner's Rule for Computing Polynomials}
\label{appendix-poly}
For the polynomial  
$p_s(x) = \sum\limits^s_{i=0} c_i x^i = c_0 + c_1 x + \cdots + c_s x^s
$
we have the \Algr \ref{alg-Polynomial} based on Horner's rule for computing this polynomial with computational complexity $\mathcal{O}(s)$.

\begin{breakablealgorithm}
\caption{Computing Polynomials by Horner's Rule}
\label{alg-Polynomial}
\begin{algorithmic}[1]
\Require The degree $s$, coefficients $c_0, c_1, \cdots, c_s$ and variable $x$ of the polynomial to be calculated. 
\Ensure $p_s(x) = \sum\limits^s_{i=0} c_i x^i$ 
\Function{Polynomial}{$c_0, \cdots, c_s, x$}
\State $T \gets c_s$;
\For{$i\in \seq{1, 2, \cdots, s}$}
\State $T \gets T \cdot x + c_{s-i}$;
\EndFor
\State \Return $T$;
\EndFunction
\end{algorithmic}
\end{breakablealgorithm}

\end{document}